\documentclass[journal]{IEEEtran}
\usepackage[utf8]{inputenc}
\usepackage{amsmath}
\usepackage{amssymb}
\usepackage{romannum}
\usepackage{etoolbox}
\usepackage{graphicx}
\usepackage{makecell}
\usepackage{booktabs}
\usepackage{ulem}
\usepackage{xcolor}
\usepackage{epstopdf}
\usepackage{verbatim}
\usepackage{hyperref}

\ifCLASSINFOpdf

\else

\fi

\begin{document}
%
\title{Grid Edge Intelligence-Assisted Model Predictive Framework for Black Start of Distribution Systems with Inverter-Based Resources}

\author{Junyuan~Zheng,~\IEEEmembership{Student~Member,}
        Salish~Maharjan,~\IEEEmembership{Member,}
        and~Zhaoyu~Wang,~\IEEEmembership{Senior~Member,~IEEE}
        \vspace{-1cm}
\thanks{The authors are with the Department of Electrical and Computer Engineering, Iowa State University, Ames, IA 50010 USA (e-mail: zhenjy@iastate.edu; salish@iastate.edu; wzy@iastate.edu).}
\thanks{Corresponding author: wzy@iastate.edu.}
}

%
%

\markboth{Journal of \LaTeX\ Class Files,~Vol.~14, No.~8, August~2015}%
{Shell \MakeLowercase{\textit{et al.}}: Grid Edge Intelligence-Assisted Model Predictive Framework for Black Start of Distribution Systems with IBRs}
%

\maketitle

\begin{abstract}
The growing proliferation of distributed energy resources (DERs) is significantly enhancing the resilience and reliability of distribution systems. However, a substantial portion of behind-the-meter (BTM) DERs is often overlooked during black start (BS) and restoration processes. Existing BS strategies that utilize grid-forming (GFM) battery energy storage systems (BESS) frequently ignore critical frequency security and synchronization constraints. To address these limitations, this paper proposes a predictive framework for bottom-up BS that leverages the flexibility of BTM DERs through Grid Edge Intelligence (GEI). A predictive model is developed for GEI to estimate multi-period flexibility ranges and track dispatch signals from the utility. A frequency-constrained BS strategy is then introduced, explicitly incorporating constraints on frequency nadir, rate-of-change-of-frequency (RoCoF), and quasi-steady-state (QSS) frequency. The framework also includes synchronizing switches to enable faster and more secure load restoration. Notably, it requires GEI devices to communicate only their flexibility ranges and the utility to send dispatch signals—without exchanging detailed asset information. The proposed framework is validated using a modified IEEE 123-bus test system, and the impact of GEI is demonstrated by comparing results across various GEI penetration scenarios.
\end{abstract}

\begin{IEEEkeywords}
Distributed energy resources, model predictive control, grid-forming, grid edge intelligence, frequency security, synchronization.
\end{IEEEkeywords}

%
\IEEEpeerreviewmaketitle

\section{Introduction}

\IEEEPARstart{T}{he} Distribution System (DS) is reportedly vulnerable to high-impact, low-probability events such as hurricanes, tornadoes, and ice storms. These events often trigger widespread blackouts that adversely impact economic and social security. Therefore, rapid load restoration following a blackout, known as black start (BS), is critical to the resilience and reliability of the power system. More importantly, BS supported by Distributed Energy Resources (DERs) and Grid Edge Intelligence (GEI) is expected to replace conventional BS resources such as diesel generators.

Black start and load-restoration strategies can generally be classified into two categories: (a) top-down and (b) bottom-up strategies \cite{Wang2015}, \cite{Wang2016}. The top-down approach relies on the transmission grid (TG) for restoration and is applicable when the blackout is localized within the DS. In the event of a large-scale power outage, typically caused by extreme weather, the bottom-up strategy helps restore the load even in the absence of an upstream transmission system. Traditionally, bottom-up strategies have been supported by BS units such as diesel generators \cite{Sun2011}. Utilities maintain diesel generators as backup units for emergency situations due to their high operating costs \cite{Patsakis2018}. However, BESS with grid-forming (GFM) inverters are now regarded as alternative BS units. The adoption of BESS is growing due to reduced production costs \cite{Wang2025} and, unlike diesel units, they can be operated during both normal and emergency periods. Additionally, the increasing prevalence of DERs, including wind turbines, Photovoltaic (PV) systems, and small-scale storage at customer premises, has provided ample non-BS resources that can be leveraged with BESS for bottom-up black start. These resources offer several advantages over diesel generators, including high self-starting speed, lower operating costs, and fast dynamic response \cite{Braun2018}.

Currently, the DERs-aided black start has emerged as an effective and economical solution for providing restoration services following a blackout \cite{liu2021}. In \cite{wang2019}, the coordination of multiple DERs is utilized to pick up the critical load during power outages. With advancements in GFM, GFM-based DERs can independently regulate system frequency and voltage, thereby enhancing their black start capability \cite{sadeque2023}, \cite{Maharjan2025}. So far, DER-aided black start strategies can be broadly classified into two categories: optimization-based \cite{Arif2018}-\cite{Arif2022} and machine learning-based strategies \cite{Bedoya2021}, \cite{Du2022}. The former approach focuses on reformulating the load restoration problem as an optimization problem to determine the final network topology, while the latter employs machine learning to train models for making load restoration decisions. 

However, most of the literature in this vein concentrates on steady-state operation conditions, while less attention has been paid to dynamic frequency security constraints during the restoration process. In practice, the frequency of a microgrid tends to decline as loads are picked up. Therefore, incorporating frequency security constraints into the restoration process is essential to ensure stable and reliable BS. In \cite{bassey2019}, the frequency response of distributed generators is derived and incorporated into load restoration. Nevertheless, the dynamic frequency behavior of the distributed generators still requires detailed simulation for a more accurate analysis. In \cite{che2019}, the frequency nadir constraint is considered during the restoration of multiple microgrids, enabling the estimation of the frequency nadir limit without simulating the complex dynamic model of BS generators. While considering only the frequency nadir constraint is insufficient, the system frequency must also satisfy the rate of change of frequency (RoCoF) and quasi-steady-state (QSS) frequency constraints during load restoration. In \cite{zhang2021}, dynamic frequency constraints are taken into account in the load restoration process, which incorporates DERs based on GFM and grid-following (GFL) inverters. Nonetheless, this approach relies on simulations, which can be computationally intensive for large-scale restoration planning problems. Hence, there is an urgent need to model dynamic frequency constraints that are both accurate and practical for load restoration. Additionally, as the microgrid frequency tends to deviate when loads are picked up, it is essential to determine whether the frequency of a formed islanded MG is synchronized with other islanded MGs and with the TG. The aforementioned studies overlook synchronization both among islanded MGs and between islanded MGs and the TG, which can result in the formation of multiple islanded MGs. Meanwhile, BESS-based BS units cannot ensure the long-term operation of restored MGs, given their finite energy capacity. Therefore, synchronization should be integrated into the blackstart process for stable load restoration.

Moreover, current bottom-up black start strategies primarily rely on utility-controlled DERs for load restoration. However, behind-the-meter (BTM) DERs, which are owned by individual customers, cannot be directly controlled by the utility due to privacy protection and ownership limitations. As a result, the potential contribution of the substantial number of BTM DERs during black start is often neglected. For instance, in \cite{song2021}, individual residential loads are directly modeled as cold loads, without accounting for the availability of BTM DERs. With the advancement of grid-side intelligence, the flexibility of behind-the-meter DERs can be estimated and incorporated into load restoration planning, accelerating the restoration process. Hence, the utilization of behind-the-meter DERs needs further investigation.

This paper proposes a model predictive framework for a bottom-up BS strategy that coordinates with multiple GEI systems to enhance load restoration. In this approach, each GEI communicates its multi-period flexibility range to the utility responsible for black start. In return, the utility provides a dispatch signal that the GEI devices track. This coordination mechanism ensures data privacy by eliminating the need for either the GEI devices or the utility to share proprietary information about their assets, network configurations, or control strategies. The utility's black start function is designed to restore the DS using multiple GFM BESS, which establish multiple microgrids. Each microgrid then coordinates with GEI to create cranking paths that bring GFL resources online. These resources help sustain the energized sections and prepare the system to restore the remaining unenergized sections. The paper also models the response of grid-forming inverters to load pickup, analyzing key frequency characteristics such as frequency nadir, rate of change of frequency, and quasi-steady-state frequency. Additionally, it incorporates the role of synchronizing switches, which enable the integration of smaller microgrids into larger ones. This integration enhances the overall power and energy capacity, facilitating faster load restoration. 

In summary, the main contributions of this paper are summarized as follows:
\begin{enumerate}
    \item A model predictive framework which enables interaction between the utility and GEI for efficient black start of DS, by sharing only the coupling variables such as flexibility range and dispatch signal. 
    \item Predictive modeling of GEI with the ability to predict flexibility range and follow utility dispatch signal. With this approach, the contribution of BTM DERs on the black start is assessed for various ratio of customers with GEI and without GEI.
    \item Predictive modeling of bottom-up BS strategy with GFM inverter's frequency security constraint, energizing and synchronizing switching constraints that enable formation and synchronization of multiple dynamic microgrids to share their energy and power capacity for faster restoration.  
\end{enumerate}


\section{Overview of the Proposed Framework}

\begin{figure*}[t] 
  \centering
  \includegraphics[scale=1]{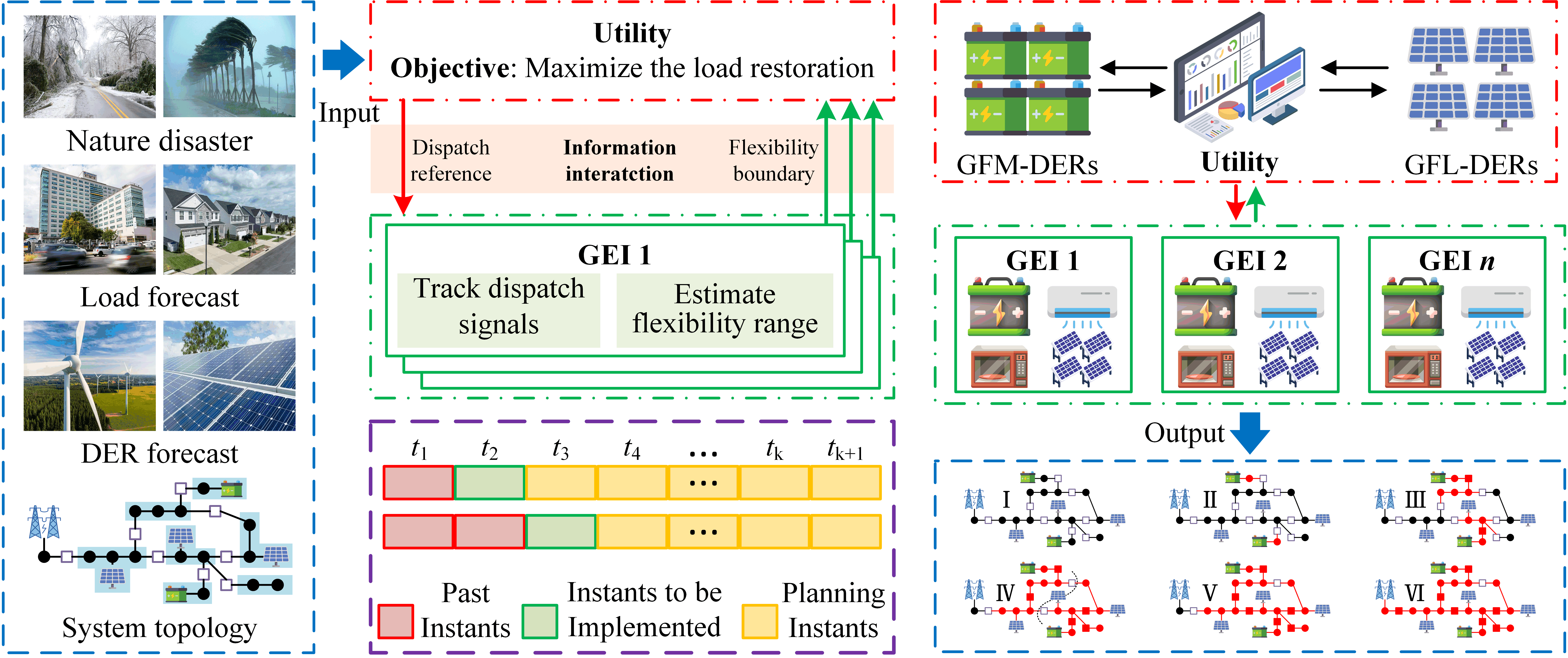} 
  \caption{Overview of the proposed framework}
  \label{the_proposed_solutionl}
\end{figure*}


As shown in Fig. \ref{the_proposed_solutionl}, GEI is deployed in residential houses equipped with BTM DERs, BES, HVAC systems, and other loads. Each GEI device is capable of estimating its flexibility range and tracking dispatch signals from the utility to optimize its operation while supporting utility functions such as black start. The GEI communicates its flexibility range to the utility, which in turn sends dispatch signals back to the GEI devices for tracking. This dynamic interaction preserves privacy, as both the GEI devices and the utility exchange only the coupling variables rather than detailed asset information. Additionally, both the GEI operations and the black start process are implemented using the principles of model predictive control (MPC).

The black start strategy follows a bottom-up approach. In this setup, multiple GFM inverters initiate the black start from different locations, forming several islanded microgrids while establishing cranking paths to bring GFL resources online. Unlike in transmission systems, load pickup is necessary during cranking path formation in distribution systems. This makes coordination between GEI devices and the utility essential to ensure dynamic frequency security throughout the restoration process.


\section{Predictive Modeling of Grid Edge Intelligence}
In this paper, residential loads are assumed to be equipped with GEI. Each GEI operates based on a predictive model that communicates its flexibility range to the utility and tracks the dispatch signal received from the utility.

The prediction horizon is defined relative to the current time $t$ as $\{t+k\Delta t: k\in\mathcal{N}_k\}$, where $\Delta t$ denotes the time step and $\mathcal{N}_k=\{1,2,\ldots,N\}$. This notation is consistently used for all predictive models, including those for GEI and the black start modeling in Section~\ref{sec_balckstart}.
\subsection{Residential Load Model with GEI} 
A residential house with GEI is assumed to be equipped with an HVAC system, a BES, PV, and other loads. The set $\mathcal{N}_h$ denotes the nodes where residential houses have GEI devices.

\subsubsection{Thermal modeling of residential house with HVAC}
\begin{figure}[t]
     \centering
     \includegraphics[scale=1]{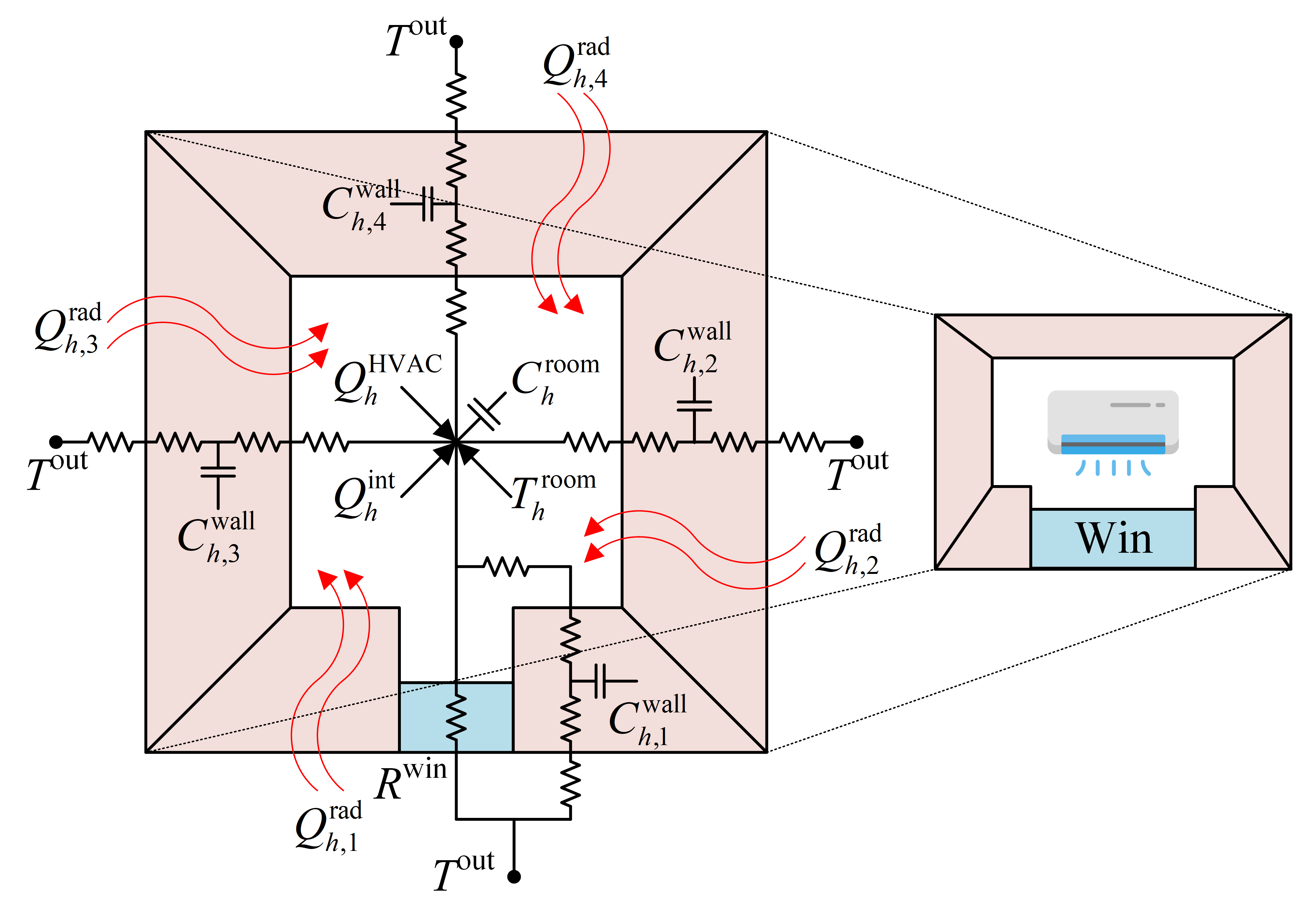}
     \caption{RC model of of a generic house.}
     \label{HVAC}
 \end{figure}

The thermal dynamics model of a generic house is illustrated in Fig.~\ref{HVAC}. The thermal model consists of five nodes, including four wall nodes and one room node. Each node is thermally connected to other nodes through thermal resistances and grounded through thermal capacitance. In this section, a typical three-resistance and two-capacitance (3R–2C) model \cite{taha2019} is employed to model its thermal dynamics. The wall temperature $T_{h,i}^{\mathrm{wall}}$ is primarily influenced by the indoor temperature $T_{h}^{\mathrm{room}}$ and the outdoor temperature $T_{h}^{\mathrm{out}}$. The temperature profile of the house $h$ with walls represented by a set $i=\{1,2,3,4\}$ can be modeled by (\ref{Wall_Thermal_Balance}), as in \cite{jin2025}: 
\begin{equation}
\begin{split}
&C_{h,i}^{\text{wall}} \frac{T_{h,t+k}^{\text{wall}} - T_{h,t+k-1}^{\text{wall}}}{\Delta t} =  \frac{T_{h,t+k}^{\text{room}} - T_{h,i,t+k}^{\text{wall}}}{R_{h,i}^{\text{wall}}} \\
& + \frac{T_{h,t+k}^{\text{out}} - T_{h,i,t+k}^{\text{wall}}}{R_{h,i}^{\text{wall}}}+ w_{h,i}^{\text{wall}} Q_{h,i,t+k}^{\text{rad,wall}}\quad \forall i \in\{1,2,3,4\}
\end{split}
\label{Wall_Thermal_Balance}
\end{equation}
Here, $C_{h,i}^{\mathrm{wall}}$ denotes the thermal capacitance and $R_{h,i}^{\mathrm{wall}}$ represents the corresponding thermal resistance. $Q_{h,i}^{\mathrm{rad,wall}}$ is the external radiative heat input to wall $i$, and $w_{h,i}^{\text{wall}}$ is a simplified weighting factor applied to $Q_{h,i}^{\mathrm{rad,wall}}$.

The indoor temperature experienced by occupants is regulated by the HVAC system. The thermal dynamics of the indoor zone can be defined by constraints~(\ref{Rule_thermal_balance_D}).
\begin{equation}
\begin{split}
&C_{h}^{\text{room}} \frac{T_{h,t+k}^{\text{room}} - T_{h,t+k-1}^{\text{room}}}{\Delta t} = \sum_{i=1}^{4} \frac{T_{h,i,t+k}^{\text{wall}} - T_{h,t+k}^{\text{room}}}{R_{h,i}^{\text{wall}}} \\
& + \frac{T_{h,t+k}^{\text{out}} - T_{h,t+k}^{\text{room}}}{R_{h,i}^{\text{win}}}+ w_{h,i}^{\text{win}} Q_{h,t+k}^{\text{rad,win}} + Q_{h,t+k}^{\text{int}} + Q_{h,t+k}^{\text{HVAC}} 
\end{split}
\label{Rule_thermal_balance_D}
\end{equation}
Here, $Q_{h}^{\mathrm{rad,win}}$ is the external radiative heat transferring into the room through the window, and $w_{h}^{\mathrm{win}}$ is the corresponding weighting factor. $Q_{h}^{\mathrm{int}}$ and $Q_{h}^{\mathrm{HVAC}}$ denote the internal heat gains and the heat supplied by the HVAC system, respectively.

\paragraph{HVAC Constraints }
The power consumed by the HVAC system is determined by the temperature difference between the HVAC setpoint temperature $T_{h}^{\mathrm{HVAC}}$ and the actual indoor temperature $T_{h}^{\mathrm{room}}$, and is expressed by constraint~\eqref{power of HVAC}, as in \cite{Maharjan2023}. Constraint \eqref{Rule_temperature_limit} defines the upper and lower bounds of the HVAC temperature, denoted by $\overline{T}_{h}^{\mathrm{HVAC}}$ and $\underline{T}_{h}^{\mathrm{HVAC}}$, respectively.
\begin{equation}
P_{h,t+k}^{\text{HVAC}} =  \left[ \frac{T_{h,t+k}^{\text{HVAC}} - T_{h,t+k}^{\text{room}}}{\text{COP}(273 + T_{h,t+k}^{\text{room}})} \right] Q_{h,t+k}^{\text{HVAC}}
\label{power of HVAC}
\end{equation}
\begin{equation}
\underline{T}_{h}^{\text{HVAC}} \leq T_{h,t+k}^{\text{HVAC}} \leq \overline{T}_{h}^{\text{HVAC}}
\label{Rule_temperature_limit}
\end{equation}

\subsubsection{BES Constraints} 
The energy storage of the BES dynamically changes with charging and discharging power and it is governed by constraint~\eqref{Rule_E}. Constraint~\eqref{Rule_E_LIMIT} defines the maximum and minimum energy storage capacities of the BES, denoted by $\overline{E}_{h}^{\mathrm{ES}}$ and $\underline{E}_{h}^{\mathrm{ES}}$, respectively. Constraints~\eqref{Rule_charge_max} and~\eqref{Rule_discharge_max} ensure that the charging power $P_{h}^{\mathrm{ES},C}$ and discharging power $P_{h}^{\mathrm{ES},D}$ do not exceed the maximum charging capacity $\overline{P}_{h}^{\mathrm{ES},C}$ and discharging capacity $\overline{P}_{h}^{\mathrm{ES},D}$, respectively. Similarly, the charging and discharging ramp limits are defined by constraints \eqref{Rule_charge_rate_max} and \eqref{Rule_discharge_rate_max}. Constraint \eqref{Rule_charge_discharge} defines that the BES cannot be charged and discharged simultaneously.
\begin{equation}
E_{h,t+k}^{\mathrm{ES}} = E_{h,t+k-1}^{\mathrm{ES}} 
+ \Delta t \left( 
\eta_{h}^{\mathrm{ES},\mathrm{C}} P_{h,t+k}^{\mathrm{ES},\mathrm{C}} 
- \frac{P_{h,t+k}^{\mathrm{ES},\mathrm{D}}}{\eta_{h}^{\mathrm{ES},\mathrm{D}}} 
\right)
\label{Rule_E}
\end{equation}
\begin{equation}
\underline{E}_{h}^{\mathrm{ES}} \leq E_{h,t+k}^{\mathrm{ES}} \leq \overline{E}_{h}^{\mathrm{ES}}
\label{Rule_E_LIMIT}
\end{equation}
\begin{equation}
0 \leq P_{h,t+k}^{\mathrm{ES},\mathrm{C}} \leq \beta_{h,t+k}^{\mathrm{ES},\mathrm{C}} \, \overline{P}_{h}^{\mathrm{ES},\mathrm{C}}
\label{Rule_charge_max}
\end{equation}
\begin{equation}
0 \leq P_{h,t+k}^{\mathrm{ES},\mathrm{D}} \leq \beta_{h,t+k}^{\mathrm{ES},\mathrm{D}} \, \overline{P}_{h}^{\mathrm{ES},\mathrm{D}}
\label{Rule_discharge_max}
\end{equation}
\begin{equation}
\underline{\Delta P}_{h}^{\mathrm{ES},\mathrm{C}} \leq \Delta P_{h,t+k}^{\mathrm{ES},\mathrm{C}} \leq \overline{\Delta P}_{h}^{\mathrm{ES},\mathrm{C}}
\label{Rule_charge_rate_max}
\end{equation}
\begin{equation}
\underline{\Delta P}_{h}^{\mathrm{ES},\mathrm{D}} \leq \Delta P_{h,t+k}^{\mathrm{ES},\mathrm{D}} \leq \overline{\Delta P}_{h}^{\mathrm{ES},\mathrm{D}}
\label{Rule_discharge_rate_max}
\end{equation}
\begin{equation}
\beta_{h,t+k}^{\mathrm{ES},\mathrm{C}} + \beta_{h,t+k}^{\mathrm{ES},\mathrm{D}} \leq 1
\label{Rule_charge_discharge}
\end{equation}
Here, $\eta_{h}^{\mathrm{ES},\mathrm{C}}$ and $\eta_{h}^{\mathrm{ES},\mathrm{D}}$ are BES charging and discharging efficiencies. $\underline{\Delta P}_{h}^{\mathrm{ES},\mathrm{C}} / \underline{\Delta P}_{h}^{\mathrm{ES},\mathrm{D}}$ and $\overline{\Delta P}_{h}^{\mathrm{ES},\mathrm{C}} / \overline{\Delta P}_{h}^{\mathrm{ES},\mathrm{D}}$ are minimum and maximum charging/discharging rates. $\beta_{h,t+k}^{\mathrm{ES},\mathrm{C}}$ and $\beta_{h,t+k}^{\mathrm{ES},\mathrm{C}}$ are binary variables.

\subsubsection{PV Constraints}
The PV is modeled using multi-time-step forecasted values $\tilde{P}_{h,t+k}^{\mathrm{PV}}$ over the prediction horizon. Accordingly, the PV output active power is defined as:
\begin{equation}
P_{h,t+k}^{\mathrm{PV}} = \tilde{P}_{h,t+k}^{\mathrm{PV}} \, x_{h,t+k}^{\mathrm{PV}}
\label{Rule_PV}
\end{equation}
where $x_{h,t+k}^{\mathrm{PV}}$ is an adjustable scaling factor constrained by $0.5 \leq x_{h,t+k}^{\mathrm{PV}} \leq 1$.

\subsubsection{Load Constraints}
The load is also modeled using predicted values $\tilde{P}_{h,t+k}^{\mathrm{load}}$ along the prediction horizon and is defined as:
\begin{equation}
P_{h,t+k}^{\mathrm{load}} = \tilde{P}_{h,t+k}^{\mathrm{load}} \, x_{h,t+k}^{\mathrm{load}}
\label{Rule_load}
\end{equation}
where $x_{h,t+k}^{\mathrm{load}}$ is an adjustable scaling factor constrained by $0.5 \leq x_{h,t+k}^{\mathrm{load}} \leq 1$.

\subsection{GEI Functions}\label{sec_GEI_formulations}
This section primarily focuses on two key functions of GEI: (a) flexibility range estimation and (b) energy optimization during both standalone and grid-connected operations. Other functions, such as forecasting and coordination, are illustrated in Fig.~\ref{flowchart}.
\subsubsection{Flexibility Range Estimation} 
GEI analyzes the power consumption of all BTM DERs, loads, and overall power consumption of the house. GEI estimates the net power consumption of a house $h$ as:
\begin{equation}
\begin{aligned}
{P}_{h,t+k}^{\text{GEI}} &=  P_{h,t+k}^{\text{ES,C}} - P_{h,t+k}^{\text{ES,D}} - P_{h,t+k}^{\text{PV}}  + P_{h,t+k}^{\text{load}}
+ P_{h,t+k}^{\text{HVAC}}
\end{aligned}
\label{lGEI POWER}
\end{equation}

The flexibility range is defined by the upper bound $\overline{P}_{h,t+k}^{\text{GEI}}$ and the lower bound $\underline{P}_{h,t+k}^{\text{GEI}}$. Along the prediction horizon, it can be obtained by solving the optimization problems defined in \eqref{upper bound} and (\ref{lower bound}).
\begin{align}
\overline{P}_{h,t+k}^{\text{GEI}} &= \arg\max \sum_{k \in \mathcal{N}_t} P_{h,t+k}^{\text{GEI}} \notag \\
\text{s.t.} \quad & (\ref{Wall_Thermal_Balance})\sim(\ref{Rule_load})
\label{upper bound}
\end{align}
\begin{align}
\underline{P}_{h,t+k}^{\text{GEI}} &= \arg\min \sum_{k \in \mathcal{N}_t} P_{h,t+k}^{\text{GEI}} \notag \\
\text{s.t.} \quad & (\ref{Wall_Thermal_Balance})\sim(\ref{Rule_load})
\label{lower bound}
\end{align}

\subsubsection{Energy Optimization}
\begin{figure}[t]
     \centering
     \includegraphics[scale=0.7]{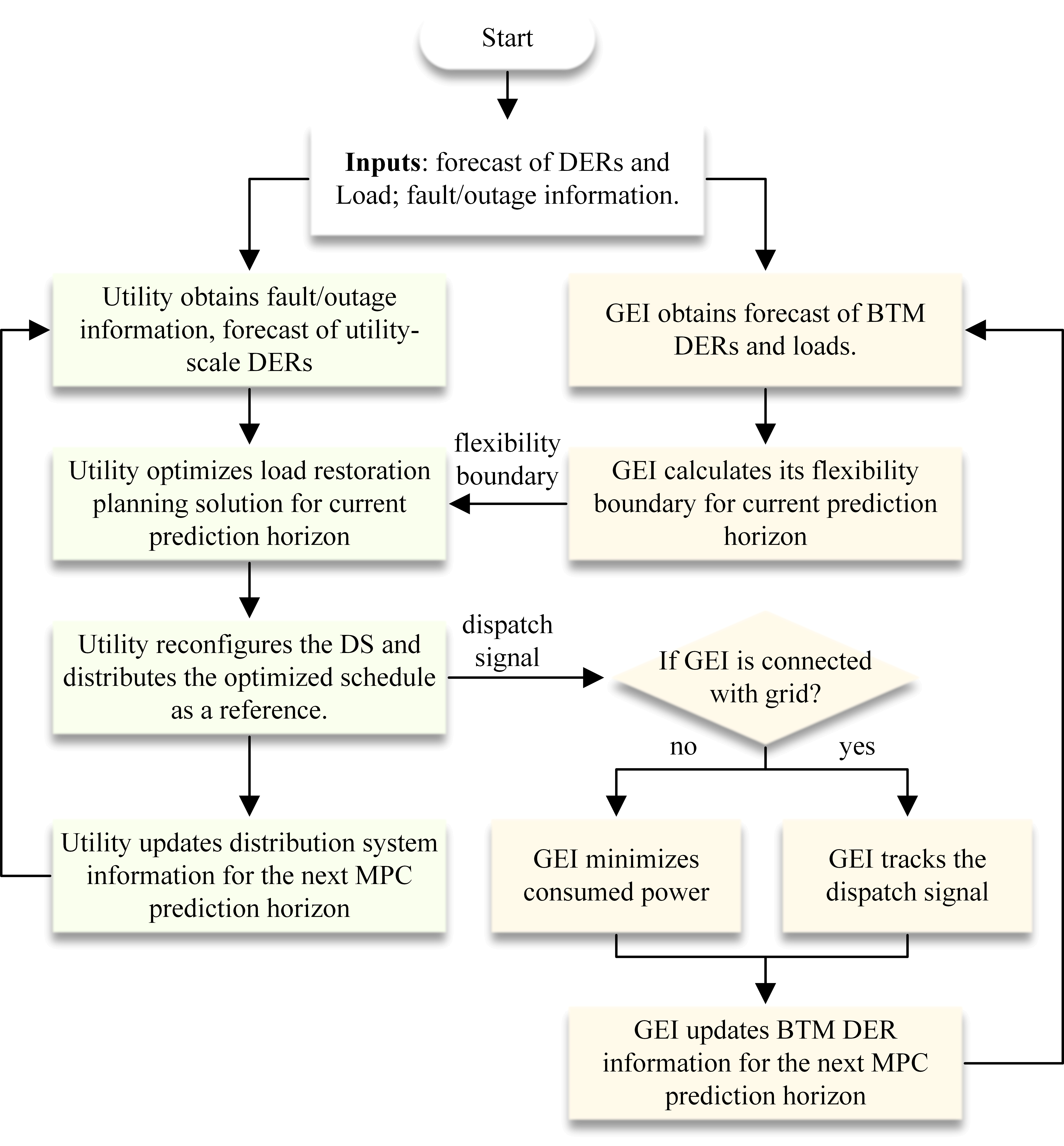}
     \caption{GEI and utility functions and their proposed coordination.}
     \label{flowchart}
 \end{figure}
The GEI optimizes the energy consumption during standalone operation, whereas it tracks the dispatch signal ($P_{h,t+k}^{\text{Ref}}$) when connected with the grid. We define the grid connection status by a binary variable $\beta^{GEI}$, to express the generic objective of GEI as: 
\begin{equation}
\begin{aligned}
\min \sum_{k \in \mathcal{N}_k} \Big[ &(1 - \beta_h^{\text{GEI}}) P_{h,t+k}^{\text{GEI}} \\
&+ \beta_h^{\text{GEI}} \left\| P_{h,t+k}^{\text{GEI}} - P_{h,t+k}^{\text{Ref}} \right\|^2 \Big] \\
\text{s.t.} \quad & (\ref{Wall_Thermal_Balance})\sim(\ref{Rule_load})
\end{aligned}
\label{GEI_model}
\end{equation}
Here, when $\beta_{h}^{\text{GEI}} = 1$, the GEI is grid connected and follows the utility dispatch signal; and when $\beta_{h}^{\text{GEI}} = 0$, the GEI continues to operate in a consumption-minimizing mode.

\subsection{GEI-Utility Coordination}
The proposed dynamic interaction between the utility and GEI is illustrated in Fig.~\ref{flowchart}. This coordination allows the utility and numerous GEI devices to collaboratively support the restoration process by exchanging only flexibility ranges and dispatch signals. As a result, neither the utility nor the GEI devices are required to share detailed asset information with one another.

The process begins with the utility gathering fault and outage data, along with forecasts for DER generation and load demand. At the same time, each GEI estimates its flexibility range using forecasted DER output and load consumption. This flexibility range is then communicated to the utility. Based on the received flexibility ranges, the utility optimizes the restoration plan, reconfigures the distribution system, and issues dispatch signals to the GEI devices.

Each GEI follows the assigned dispatch signal when connected to the utility. If a GEI is not yet connected during restoration, it operates in a standalone mode, minimizing its energy consumption. Finally, each GEI updates its BTM DER information in preparation for the next prediction horizon.
 
\section{Predictive Modeling Black Start of Distribution System} \label{sec_balckstart}
This section presents a predictive model for the bottom-up black start strategy, incorporating frequency security constraints, switching constraints, and synchronization switching constraints. The black start model is then formulated as a mixed-integer problem to maximize load restoration.

\subsection{Formulation}
The objective is to maximize the total load restoration over the $\mathcal{N}_k$ for each time $t$, which is formulated as:
\begin{align}
\max &\sum\limits_{k \in \mathcal{N}_k} \sum_{i \in \mathcal{N}_i} \sum_{p \in \mathcal{N}_p}P^D_{i,p,k} \Delta t\\
\text{s.t.} \quad & (\ref{Drop_voltage})\sim(\ref{TG_frequency}) \nonumber
\end{align}

Here, $\mathcal{N}_i$ and $\mathcal{N}_p$ represent set of load buses and phases. $P^D$ denotes active power restored.

\subsection{GFM-BESS Constraints}
\begin{figure}[t]
     \centering
     \includegraphics[scale=1]{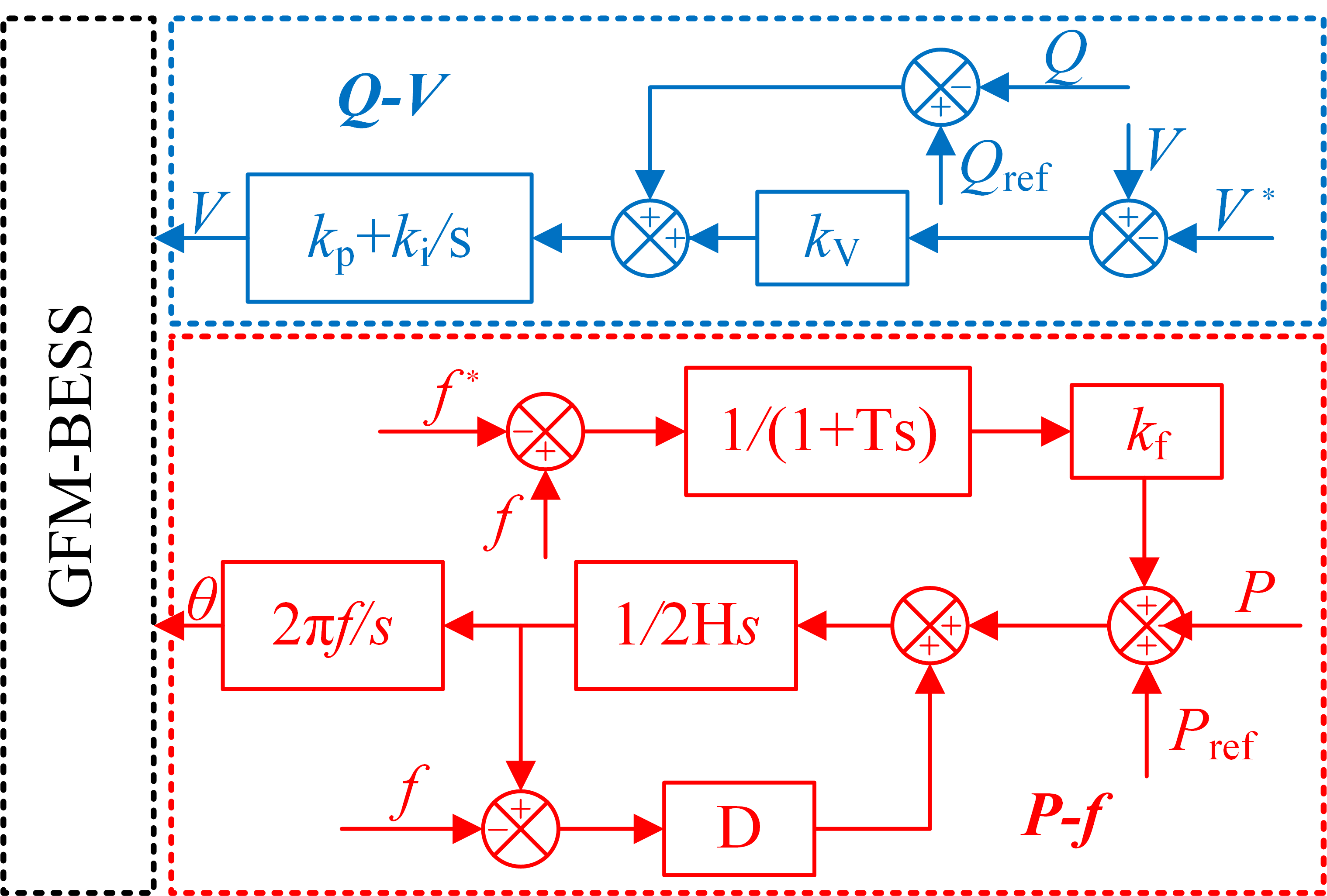}
     \caption{VSG-based control of GFMI}
     \label{GFMI}
 \end{figure}

The typical structure of the virtual synchronous generator (VSG) control is illustrated in Fig.~\ref{GFMI}. VSG control is a grid-forming strategy that emulates the dynamic behavior of a synchronous generator, providing both frequency and voltage support. In this work, all GFM-BESS are equipped with VSG control. The set $\mathcal{N}_R$ denotes the location of nodes with GFM-BESS. To ensure compatibility with the branch flow model, a new voltage variable is introduced as \( v = \lvert V \rvert^2 \), where \( \lvert V \rvert \) denotes the voltage magnitude.

The voltage magnitude deviation $\Delta v_{i,p,k}$ and frequency deviation $\Delta f_{i,k}^{QSS}$ of the GFM-BESS at QSS are defined by constraints \eqref{Drop_voltage} and \eqref{Drop_frequency}, as described in \cite{Maharjan2025}:
\begin{equation}
v_{i,p,t+k} = (V^*)^2 + \Delta v^{cc}_{i,t+k}
\label{Drop_voltage}
\end{equation}
\begin{equation}
\Delta f_{i,t+k}^{QSS} = f^*\frac{\sum_{p \in \mathcal{N}_{P}} \Delta P_{i,p,t+k}^{ES}}{S^{\text{rat}}_i (D_i + k^f_i)}
\label{Drop_frequency}
\end{equation}

The upper and lower bounds for these deviations are constrained by \eqref{Voltage_boundary} and \eqref{Frequency_boundary}, respectively:
\begin{equation}
-0.1025(V^*)^2 \leq \Delta v_{i,p,k} \leq 0.1025(V^*)^2
\label{Voltage_boundary}
\end{equation}
\begin{equation}
\Delta \underline{f}_{i}^{QSS} \leq \Delta f_{i,t+k}^{QSS} \leq \Delta \overline{f}_{i}^{QSS}
\label{Frequency_boundary}
\end{equation}
where $V^*$ and $f^*$ represent the nominal voltage and frequency references, respectively. The $S^{\text{rat}}$ is the rated capacity of the inverter. $\Delta \underline{f}_{i}^{QSS}$ and $\Delta \overline{f}_{i}^{QSS}$ denote the minimum and maximum frequency deviations at QSS.

In addition, the dynamic energy storage behavior of the GFM-BESS is modeled by constraint \eqref{battery ch_dis}. The active power $P_{i,p}^{ES}$ and reactive power $Q_{i,p}^{ES}$ of the inverter are limited by the rated capacity, as shown in constraint \eqref{Power_capacity_limit}:
\begin{equation}
E_{i,t+k}^{ES} = E_{i,t+k-1}^{ES} + \Delta t \sum_{p \in \mathcal{N}_{P}} \Delta P_{i,p,t+k}^{PE}
\label{battery ch_dis}
\end{equation}
\begin{equation}
\max_{p \in \mathcal{N}_{P}} \left\{(P_{i,p,t+k}^{ES})^2 + (Q_{i,p,t+k}^{ES})^2\right\} \leq \left(\frac{1}{3} S^{\text{rat}}_i\right)^2
\label{Power_capacity_limit}
\end{equation}

\subsubsection{Frequency Security Constraints}

To model the frequency response of GFM-BESS during load restoration, constraints \eqref{RoCoF_frequency} and \eqref{nad_frequency} are employed to calculate the RoCoF $f^{RoCoF}_{i,k}$ and the frequency nadir deviation $\Delta f^{\text{nad}}_{i,k}$ at QSS, respectively:
\begin{equation}
f^{RoCoF}_{i,t+k} = \frac{\sum_{p \in \mathcal{N}_{P}} \Delta P_{i,p,t+k}^{ES}}{2 S_i^{\text{rat}} H_i}
\label{RoCoF_frequency}
\end{equation}
\begin{equation}
\Delta f^{\text{nad}}_{i,t+k} = f^*\frac{\sum_{p \in \mathcal{N}_{P}} \Delta P_{i,p,t+k}^{ES}}{S_i^{\text{rat}} (D_i + k_i^f)} (1 + \gamma_i) 
\label{nad_frequency}   
\end{equation}

Constraints \eqref{rocof_frequency_bound} and \eqref{nad_frequency_bound} define the boundaries for the RoCoF and frequency nadir deviations of the GFM-BESS, respectively:
\begin{equation}
\underline{f}^{RoCoF}_{i} \leq f^{RoCoF}_{i,t+k} \leq \overline{f}^{RoCoF}_{i} 
\label{rocof_frequency_bound}
\end{equation}
\begin{equation}
\Delta \underline{f}_{i}^{\text{nad}} \leq \Delta f^{\text{nad}}_{i,t+k} \leq \Delta \overline{f}_{i}^{\text{nad}} 
\label{nad_frequency_bound}
\end{equation}
where $\underline{f}^{RoCoF}_{i}$/$\overline{f}^{RoCoF}_{i}$ and $\Delta \underline{f}_{i}^{\text{nad}}$/$\Delta \overline{f}_{i}^{\text{nad}}$ represent the lower and upper bounds of RoCoF and frequency nadir deviations, respectively.

\subsubsection{Synchronization constraints}
During the synchronization period, the frequency among different GFM-BESS can be aligned by adjusting their reference frequency, represented by $\Delta f^*$. Constraint \eqref{synchronization} allows for frequency penetration in GFM-BESS units attempting to synchronize with other GFM-BESS, while constraint \eqref{syn_frequency_bound} defines the acceptable boundaries for this frequency penetration.
\begin{equation}
f_{i,t+k} =  f^{*}\left(1 - \frac{\sum_{p \in \mathcal{N}_{P}} P_{i,p,t+k}^{ES}}{S_{i}^{\text{rat}} (D_i + k_{i}^{f})}\right)+\Delta f_{i,t+k}^{*}\sum_{b \in \mathcal{N}_{R}} \delta_{b,t+k}
\label{synchronization}
\end{equation}
\begin{equation}
\Delta \underline{f}_{i}^{\text{syn}} \leq \Delta f_{i,t+k}^{*} \leq \Delta \overline{f}_{i}^{\text{syn}} 
\label{syn_frequency_bound}
\end{equation}
where $\delta_{b}$ is a binary variable and $\Delta f_{i}^{*}$ represents frequency penetration. $\delta_{b}$ is set as 1 only during the synchronization period. $\Delta \underline{f}_{i}^{\text{syn}}$ and $\Delta \overline{f}_{i}^{\text{syn}}$ are the lower and upper bounds of frequency penetration, respectively.

\subsection{Switching Constraints}

DS generally has multiple bus blocks, as shown in Fig. \ref{POWER SYSTEM123}, interconnected by two types of switches: Energizing Switches (ESWs) and Synchronizing Switches (SSWs). ESWs are used to energize inactive sections of the distribution system, while SSWs ensure frequency synchronization when interconnecting two islanded microgrids. We define all ESWs by the set $\mathcal{W}_{\text{ESW}}$ and all SSWs by the set $\mathcal{W}_{\text{SSW}}$. All switches are modeled by a negligible-impedance line $(i,j)$ where $i, j \in \mathcal{N}$.

\subsubsection{Energizing switch} We define the active/inactive status of ESW by a binary variable $y^L$. The switching conditions for all ESWs $(i,j) \in \mathcal{W}_{\text{ESW}}$ can be defined as:
\begin{equation}
y^{L}_{ij,t+k} \leq y^{B}_{i,t+k-1} + y^{B}_{j,t+k-1}
\label{ESW_1}
\end{equation}
\begin{equation}
0 \leq \Delta y^{L}_{ij,t+k} \leq 2 - y^{B}_{i,t+k-1} - y^{B}_{j,t+k-1}
\label{ESW_2}
\end{equation}

A constraint \eqref{ESW_1} ensures that an ESW can only close if at least one of the buses is in an active state defined by $y^B$. Another constraint \eqref{ESW_2} prevents the ESW from closing if both connected buses are in an active state.

\subsubsection{Synchronizing switch} The switching constraint for ESWs can be defined for all $(i, j) \in \mathcal{W}_{\text{ESW}}$ as:
\begin{equation}
y^{L}_{ij,t+k} \leq y^{B}_{i,t+k-1} + y^{B}_{j,t+k-1}
\label{SSW_0}
\end{equation}
\begin{equation}
z^{L}_{ij,t+k} = \Delta y^{L}_{ij,t+k} (y^{B}_{i,t+k-1} + y^{B}_{j,t+k-1} - y^{L}_{ij,t+k})
\label{SSW_1}
\end{equation}
\begin{equation}
-(1 - z^{
L
}_{ij,t+k}) M - \epsilon \leq P_{ij,t+k} \leq \epsilon + (1 - z^{L}_{ij,t+k}) M
\label{SSW_P}
\end{equation}
\begin{equation}
-(1 - z^{L}_{ij,t+k}) M - \epsilon \leq Q_{ij,t+k}  \leq \epsilon + (1 - z^{L}_{ij,t+k}) M
\label{SSW_Q}
\end{equation}
\begin{equation}
-(1 - y_{ij,t+k}^{L})M - \epsilon \leq f_{i,t+k} - f_{j,t+k} \leq \epsilon + (1 - y_{ij,t+k}^{L})M
\label{SSW_F}
\end{equation}
where $M$ and $\epsilon$ are big and small numbers. Constraint \eqref{SSW_0} ensures that an SSW can be close even when the connecting buses are both active. Constraint \eqref{SSW_1} defines the instant when the SSW is closed. Constraints \eqref{SSW_P} and \eqref{SSW_Q} ensure that the power flow through the SSW is zero during the instant of closing the SSW to ensure that the voltage and angle of the connecting buses are sufficiently close. Constraint \eqref{SSW_F} ensures that the frequencies of the two islanded zones are matched prior to interconnection.

\subsection{Bus Blocks Energizing Constraints}

The set of bus blocks is defined by $\mathcal{B}$, and the switches associated with each bus block $b \in \mathcal{B}$ are defined by the set $\mathcal{W}_{b}$. The energizing constraints for each bus block are modeled as follows.
\begin{equation}
y^{BB}_{b,t+k} \geq y^L_{ij,t+k},\forall \{ \, b \in \mathcal{B} \;\big|\, (i, j) \in \mathcal{W}_b\}
\label{Bus_Block_1}
\end{equation}
\begin{equation}
y^{BB}_{b,t+k} \geq y^{BB}_{b,t+k-1},\forall \, b \in \mathcal{B} 
\label{Bus_Block_2}
\end{equation}
\begin{equation}
y_{ij,t+k}^{L} = y_{b,t+k}^{BB} ,\forall\{\, b \in \mathcal{B} \;\big|\, (i, j) \in \mathcal{W}_b\}
\label{Bus_Block_3}
\end{equation}
\begin{equation}
y_{i,t+k}^{B} = y_{b,t+k}^{BB} ,\forall \{ \, b \in \mathcal{B} \;\big|\, (i) \in \mathcal{L}_b\}
\label{Bus_Block_4}
\end{equation}
\begin{equation}
\sum_{(i,j) \in \mathcal{W}_b} y_{ij,t+k}^{L} - \sum_{(i,j) \in \mathcal{W}_b} y_{ij,t+k-1}^{L} 
\leq M y_{b,t+k}^{BB} + 1
\label{Bus_Block_5}
\end{equation}
Here, $\mathcal{L}_{\text{b}}$ and $\mathcal{N}_{\text{b}}$ are the set of lines and nodes associated with bus block \textit{b}. Constraint \eqref{Bus_Block_1} ensures that a bus block becomes active only if at least one switch connected to it has been activated. Constraint \eqref{Bus_Block_2} guarantees that once a bus block is energized, it remains energized in all subsequent time steps. Constraint \eqref{Bus_Block_3} ensures that energizing a bus block activates all lines within that block. Constraint \eqref{Bus_Block_4} ensures that energizing a bus block activates all buses within that block. Constraint \eqref{Bus_Block_5} restricts the number of switches that can be activated at any single time step for each bus block, specifically differentiating between active and inactive bus blocks.

\subsection{Linear power flow model adaptive to network topology}
A linearized model for a three-phase unbalanced system, as proposed in \cite{cheng2022}, is leveraged to develop a switching-enabled power flow formulation.
\subsubsection{Power constraints} Constraints~\eqref{line_power_1} and~\eqref{line_power_2} impose big-M constraints on the active and reactive power flows (\(P_{ij,t}\) and \(Q_{ij,t}\)) of line \(ij\), ensuring that power flow is allowed only when the line is active, as indicated by the binary variable \(y_{ij,t}^{L}\). Constraints~\eqref{line_power_flow_1} and~\eqref{line_power_flow_2} define the nodal power balance.
\begin{equation}
- M^{\mathcal{N}_{ip}} y_{ij,t}^{L} \leq P_{ij,t} \leq M^{\mathcal{N}_{ip}} y_{ij,t}^{L}
\label{line_power_1}
\end{equation}
\begin{equation}
- M^{\mathcal{N}_{ip}} y_{ij,t}^{L} \leq Q_{ij,t} \leq M^{\mathcal{N}_{ip}} y_{ij,t}^{L}
\label{line_power_2}
\end{equation}
\begin{equation}
P_{ij,k} = \sum_{l \in \mathcal{N}_{j}} P_{jl,k} + p_{j,k}
\label{line_power_flow_1}
\end{equation}
\begin{equation}
Q_{ij,k} = \sum_{l \in \mathcal{N}_{j}} Q_{jl,k} + q_{j,k}
\label{line_power_flow_2}
\end{equation}
\subsubsection{Voltage constraints} Constraint~\eqref{line_voltage_3} ensures that the nodal voltage  of energized nodes is within the safe range. Constraints~\eqref{Power_voltage_1} and ~\eqref{Power_voltage_2} describe the voltage difference between two end nodes of each connected line.
\begin{equation}
0.9025^{\mathcal{N}_{ip}} y_{j,t}^{B} \leq v_{j,t} \leq 1.1025^{\mathcal{N}_{ip}} y_{j,t}^{B}
\label{line_voltage_3}
\end{equation}
\begin{equation}
v_{j,t} \leq v_{i,t}^{\mathcal{N}_{ip}} - 2\left(\overline{R}_{ij} P_{ij,t} + \overline{X}_{ij} Q_{ij,t} \right) 
+ M^{\mathcal{N}_{ip}} (1 - y_{ij,t}^{L})
\label{Power_voltage_1}
\end{equation}
\begin{equation}
v_{j,t} \geq v_{i,t}^{\mathcal{N}_{ip}} - 2\left(\overline{R}_{ij} P_{ij,t} + \overline{X}_{ij} Q_{ij,t} \right) 
- M^{\mathcal{N}_{ip}} (1 - y_{ij,t}^{L})
\label{Power_voltage_2}
\end{equation}
where $P_{ij}$ and $Q_{ij}$ are the active and reactive power on line $(i,j)$. M is a big number for the relaxation of power flow constraints. $\overline{R}{ij}$ and $\overline{X}{ij}$ denote the resistance and reactance matrices of the line $(i,j)$.

\subsection{GFL inverter and GEI constraints }
We adopt GFL inverters, such as PVs, from \cite{Maharjan2025}, but do not discuss them explicitly due to space limitations. Each GEI device $h$ provides a flexibility range $(\overline{P}{h,t+k}^{\text{GEI}}, \underline{P}{h,t+k}^{\text{GEI}})$ to the utility, which introduces a new constraint in the proposed framework. This constraint is defined for all $h \in \mathcal{N}_h$ as:

\begin{align}
    \overline{P}_{h,t+k}^{\text{GEI}}\le p^{dis}_{h,t+k} \le \underline{P}_{h,t+k}^{\text{GEI}}
\end{align}
Here, $p^{dis}$ is a dispatch signal communicated to GEI device. 
\subsection{Modeling transmission grid outage and recovery}
During the BS process, DS is disconnected from an out-of-service TG and subsequently reconnected once the TG has been restored. The constraints related to the TG are modeled as follows:
\begin{equation}
(P^{TG}_{g,t+k})^2 + (Q^{TG}_{g,t+k})^2 \leq (SS^{rat})^2
\label{TG_power}
\end{equation}
\begin{equation}
1.0 y^{TG}_{g,t+k} \geq v_{g,t+k} \geq 1.0 y^{TG}_{g,t+k}
\label{TG_voltage}
\end{equation}
\begin{equation}
60 y^{TG}_{g,t+k} \geq f_{g,t+k} \geq 60 y^{TG}_{g,t+k}
\label{TG_frequency}
\end{equation}

Constraint \eqref{TG_power} defines the thermal power limit of the distribution substation. Constraints \eqref{TG_voltage} and \eqref{TG_frequency} define the voltage and frequency values of the TG, respectively. By parameterizing $y^{TG}_{g}$, the active/inactive status of TG can be simulated.

\section{Results}

This section evaluates the performance of the proposed predictive framework for the GEI-assisted bottom-up black start strategy in the modified IEEE 123-bus DS. A blackout is simulated by the outage of the TG, followed by its recovery after a few hours. During this period, the proposed framework demonstrates black start and restoration by forming multiple MGs with the help of GFM and GFL inverters and the GEI devices. Detailed results are presented for a study with 100\% GEI, highlighting the prediction of flexibility ranges, interaction with the utility, and tracking of the dispatch signal. Additionally, the sequential formation of cranking paths, starting from the GFM and continuing until the energization of all bus blocks across multiple MGs, is shown to comply with dynamic frequency security limits. A comparative analysis for various levels of GEI in the IEEE feeder is also presented to highlight the contribution of GEI in system black start and restoration. The proposed formulation is of mixed-integer nature with a linear objective and a few quadratic constraints, solved using Gurobi 11.0.3.

\begin{figure}[t]
     \centering
     \includegraphics[scale=0.96]{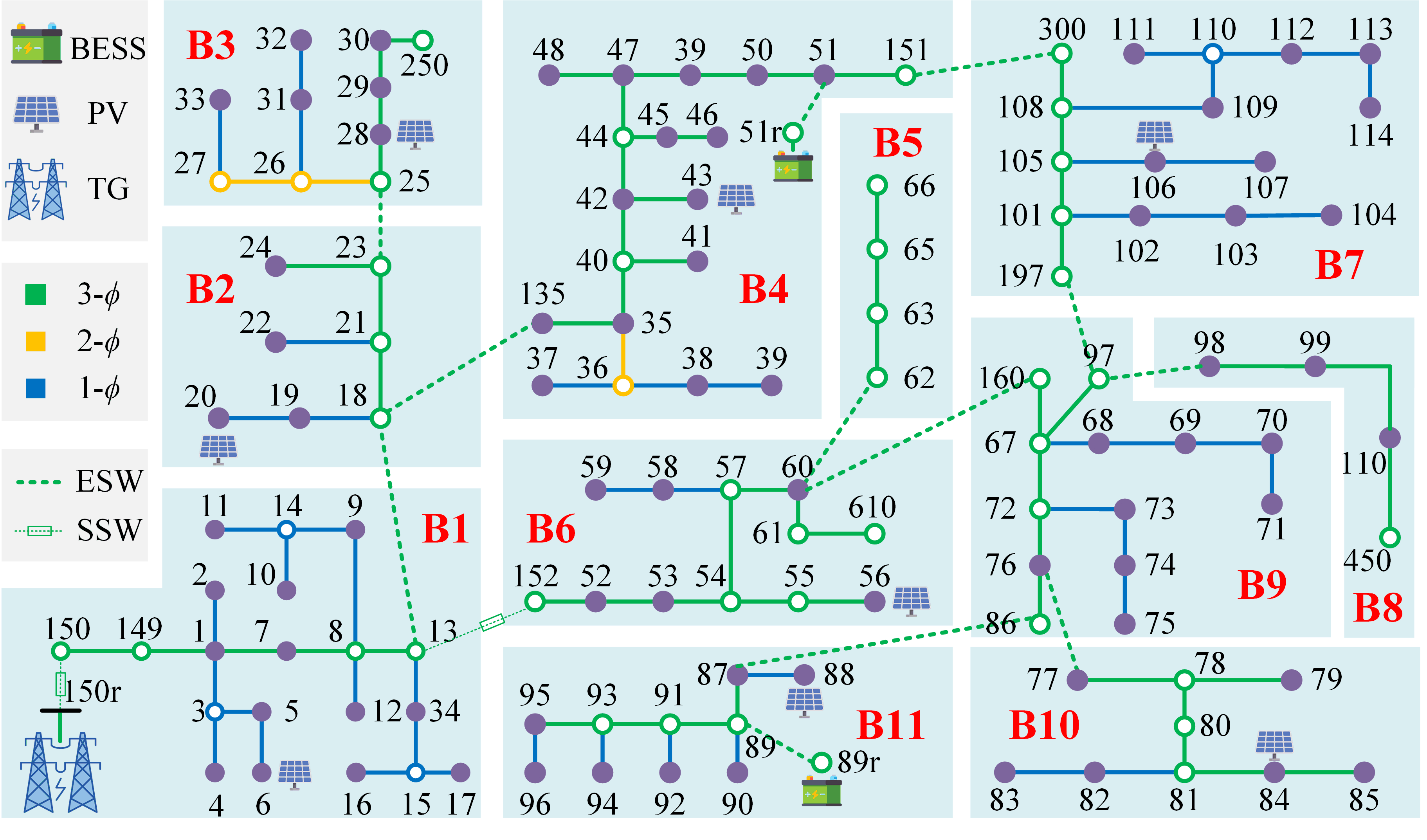}
     \caption{Modified IEEE-123 bus distribution system.}
     \label{POWER SYSTEM123}
 \end{figure}
 
\renewcommand{\arraystretch}{1.2}
\begin{table}[t]
\centering
\caption{The DERs Parameters}
\begin{tabular}{ccc|ccc}
\hline\hline
\textbf{DER} & \textbf{Capacity} & \textbf{Location} & \textbf{DER} & \textbf{Capacity} & \textbf{Location} \\
\hline
PV1   & 24.5kW  & 6    & PV2   & 44.1kW  & 20   \\
PV3   & 24.5kW  & 28   & PV4   & 24.5kW  & 43   \\
PV5   & 24.5kW  & 56   & PV6   & 12.25kW & 84   \\
PV7   & 24.5kW  & 88   & PV8   & 24.5kW  & 106  \\
BESS1 & 24.5kW  & 51r  & BESS2 & 24.5kW  & 89r  \\
\hline\hline
\end{tabular}
\label{DERs_Parameters}
\end{table}

\renewcommand{\arraystretch}{1.2}
\begin{table}[t]
\centering
\caption{The Switch Parameters}
\begin{tabular}{ccc|ccc}
\hline\hline
\textbf{Switch} & \textbf{Location} & \textbf{Type} & \textbf{Switch} & \textbf{Location} & \textbf{Type} \\
\hline
S1   & (13, 18)  & ESW  & S2   & (18, 135)  & ESW   \\
S3   & (23, 25)  & ESW   & S4   & (151, 300)  & ESW   \\
S5   & (51r, 51)  & ESW   & S6   & (60, 62) & ESW   \\
S7   & (13, 152)  & SSW   & S8   & (60, 160)  & ESW  \\
S9   & (97, 197)  & ESW  & S10  & (97, 98)  & ESW  \\
S10  & (76, 77)  & ESW  & S12  & (86, 87)  & ESW  \\
S13  & (89r, 89)  & ESW  & S14  & (150r, 150)  & SSW  \\
\hline\hline
\end{tabular}
\label{Switch_Parameters}
\end{table}

\subsection{Test System and Framework Parameters}
As depicted in Fig. \ref{POWER SYSTEM123}, the modified IEEE 123-bus distribution system consists of 88 load nodes, 8 GFL PVs, and 2 GFM BESS units. In this system, residential loads are equipped with GEI devices capable of communicating their flexibility ranges to the utility and following dispatch signals during a black start, as introduced in Section \ref{sec_GEI_formulations}. The sizes and locations of the PVs and BESS are provided in Table \ref{DERs_Parameters}. The entire system is divided into 11 bus blocks, which are interconnected through either an ESW or an SSW. The bus blocks are labeled in red text, and their structure is illustrated in Fig. \ref{POWER SYSTEM123}. The locations of the SSWs and ESWs are listed in Table \ref{Switch_Parameters}. 

Each house with GEI is equipped with a 5–7 kW rooftop PV system and a 3–5 kW HVAC unit. The BESS has a capacity ranging from 15 to 20 kWh, with an initial SoC randomly set between 40\% and 50\%. In addition, the peak load of each GEI ranges from 10 to 15 kW. These parameters are randomly selected to model all the houses with GEI.

Both the GEI and the utility's black start functions are implemented using MPC with a prediction horizon of 3 hours and a time step of 15 minutes. 

\subsection{Black start with 100\% GEI}
\subsubsection{Utility-level analysis}
\begin{figure}[t]
     \centering
     \includegraphics[scale=0.9]{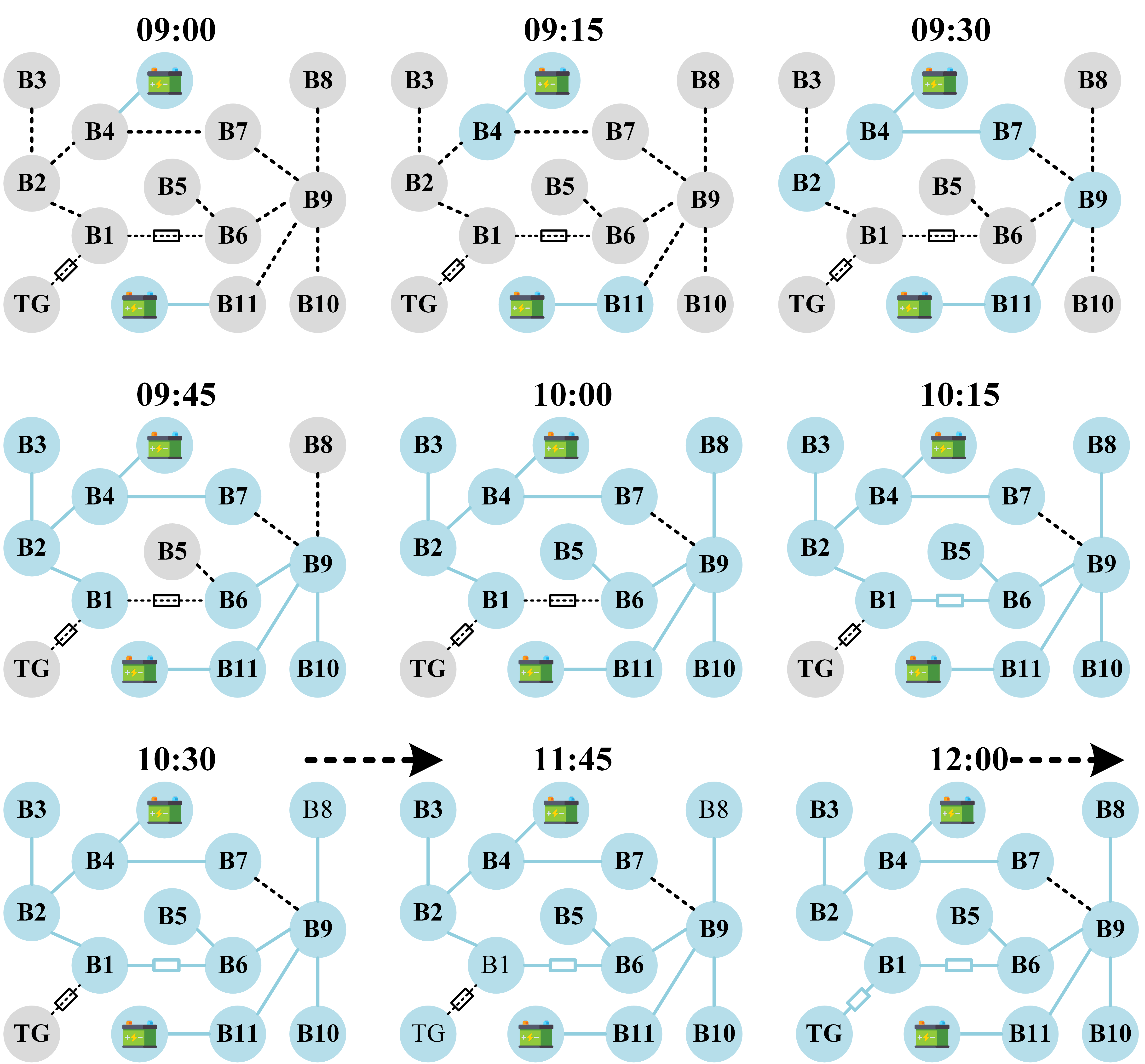}
     \caption{Sequential establishment of cranking paths in BS process.}
     \label{Load Restoration Sequence}
 \end{figure}

\begin{figure}[t]
     \centering
     \includegraphics[scale=1]{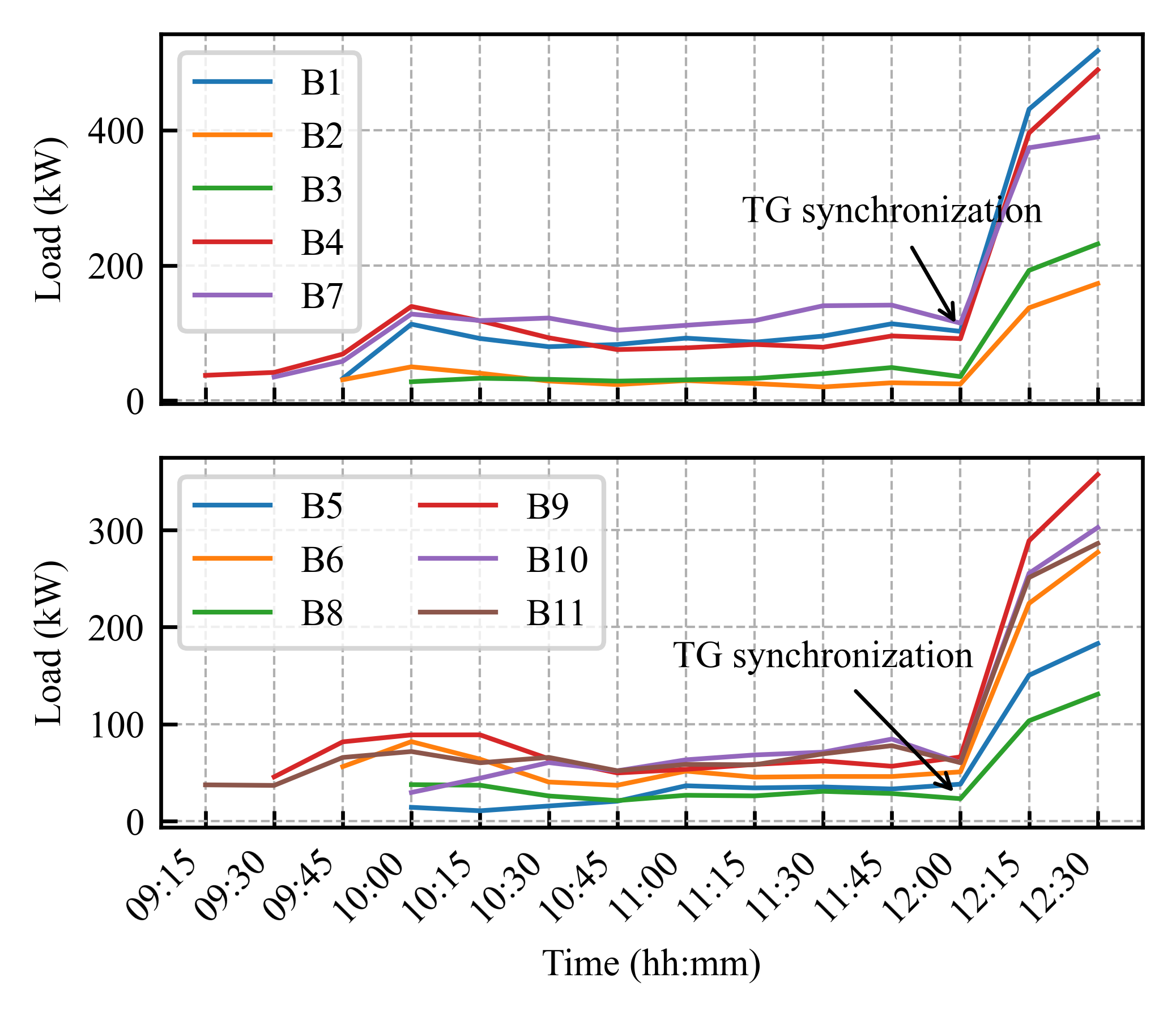}
     \caption{Restoration of loads in the proposed method.}
     \label{Restoration of loads}
 \end{figure}
The proposed framework is studied under a blackout scenario caused by TG failure, with the TG recovering at 11:45. The BS process begins at 9:00 and continues until the DS is synchronized with the TG at 12:00. The sequential establishment of cranking paths during the BS process, obtained from the proposed framework, is shown in Fig. \ref{Load Restoration Sequence}.

Initially, all bus blocks are inactive and disconnected, with all switches in the open position. The BS starts with the activation of two GFM-based BESS units at buses 51r and 89r, forming two islanded MGs. Subsequently, neighboring bus blocks are energized one after another with support from these GFM-based BESS units. For example, at 9:15, bus blocks B4 and B11 are energized by closing the ESWs adjacent to the respective BESS units. Then, at 9:30, bus blocks B2 and B7 are energized within the first MG, while bus block B9 is energized within the second MG. Both islanded MGs continue to expand their boundaries until all bus blocks are energized by 10:00. At 10:15, the SSW between bus blocks B1 and B6 is closed, merging the two islanded MGs into a unified microgrid.  At 11:45, the TG is restored but remains disconnected from the islanded DS. Finally, at 12:00, synchronization between the TG and DS completes the BS process, and the system operates continuously in this final configuration. The loads served on various bus blocks after energization are shown in Fig. \ref{Restoration of loads}. For instance, the load on bus block B5 is not served by the utility until 10:00 and is served afterward. In this case study with 100\% GEI, the load served by the utility is based on the mutual interaction between the utility’s capacity and the flexibility range of GEI devices. This interaction is discussed in a later part of this section. After TG synchronization, DS is able to serve more load than during the outage period, as seen in Fig. \ref{Restoration of loads}.

\begin{figure}[t]
     \centering
     \includegraphics[scale=1]{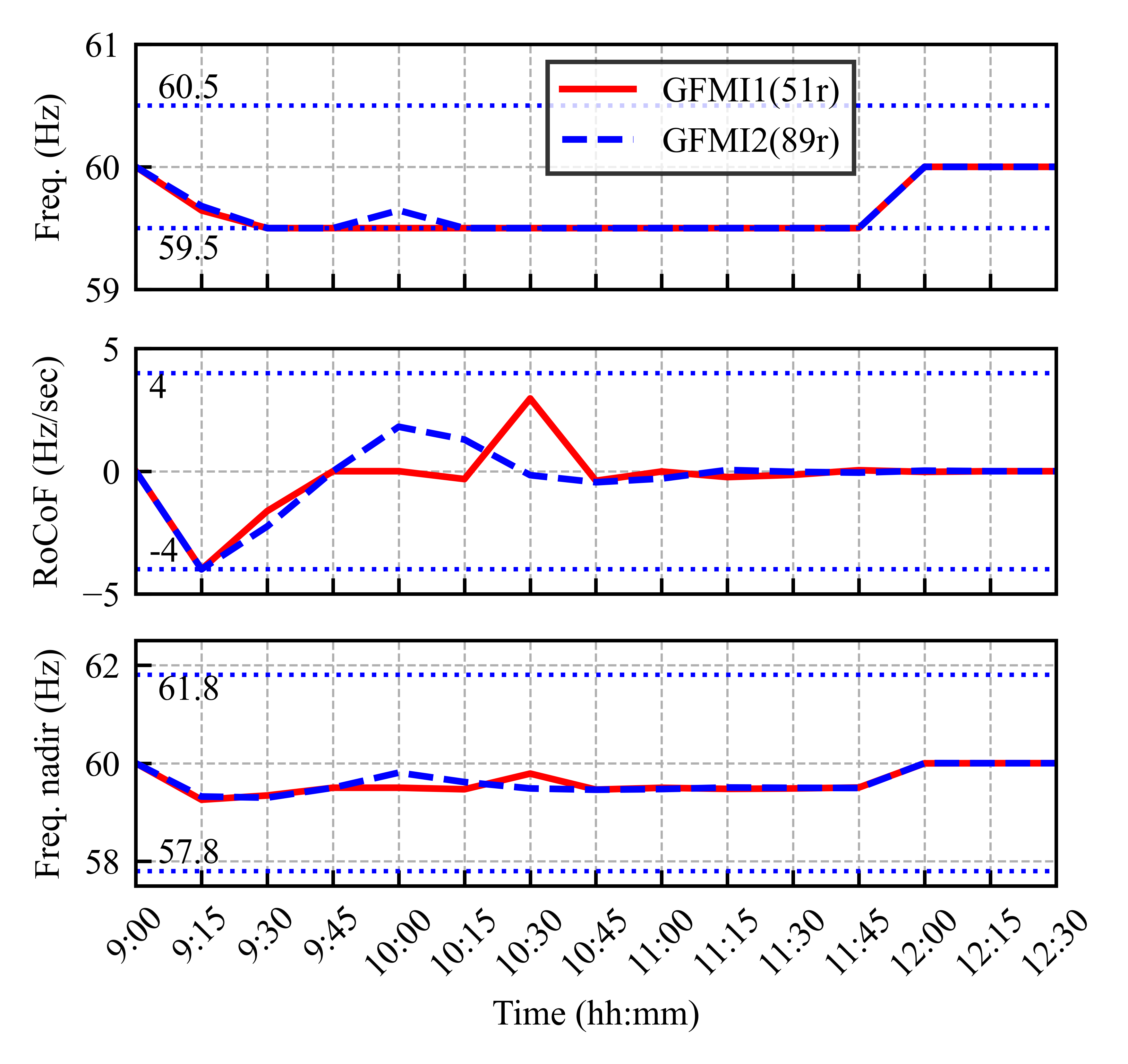}
     \caption{Frequency responses of VSG-based GFMIs.}
     \label{frequency response}
 \end{figure}

\begin{figure}[th]
     \centering
     \includegraphics[scale=1]{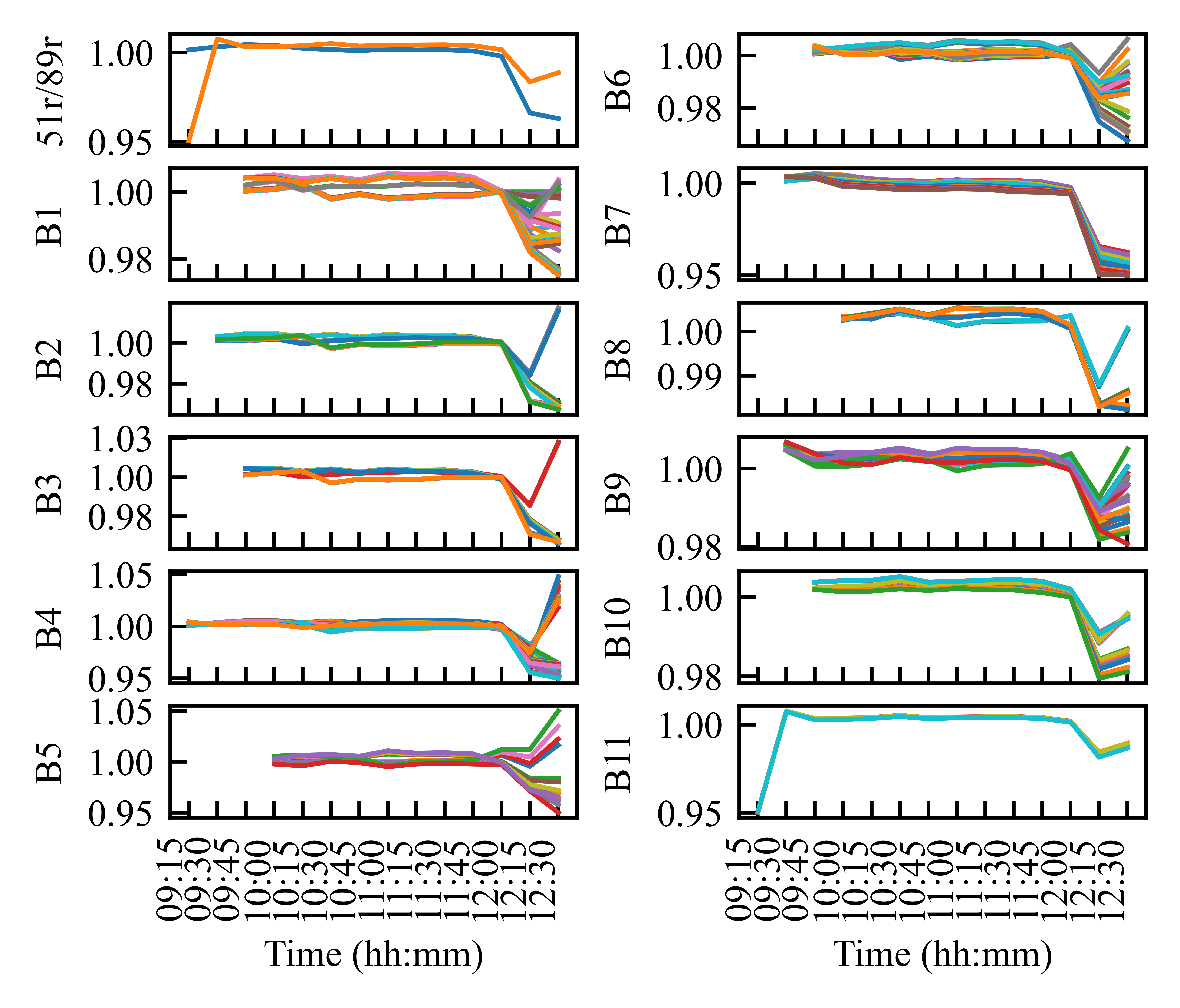}
     \caption{Nodal voltage magnitudes at bus-blocks in the proposed method.}
     \label{Nodal voltage magnitudes}
 \end{figure}

During the entire BS process, the frequency responses of the GFM-based BESS units remained within the defined frequency security constraints, as shown in Fig. \ref{frequency response}. At 9:00, when the GFM-based BESS units are started, they operate at 60 Hz since no loads are picked up at that moment. Their frequencies gradually and independently decline as they begin picking up loads, continuing until 10:00. At 10:15, the BESS units are synchronized, and their frequencies closely follow each other for the remainder of the process. Notably, their frequencies return to 60 Hz at 12:00 upon synchronization with the TG. In the same figure, both $RoCoF$ and $f^{nad}$ remain within acceptable limits, indicating a smooth establishment of cranking paths without violating dynamic frequency security constraints. In addition, the node voltages on each bus block are within 0.95 to 1.05 p.u. throughout the BS process, as shown in Fig. \ref{Nodal voltage magnitudes}.

\begin{figure}[th]
     \centering
     \includegraphics[scale=1]{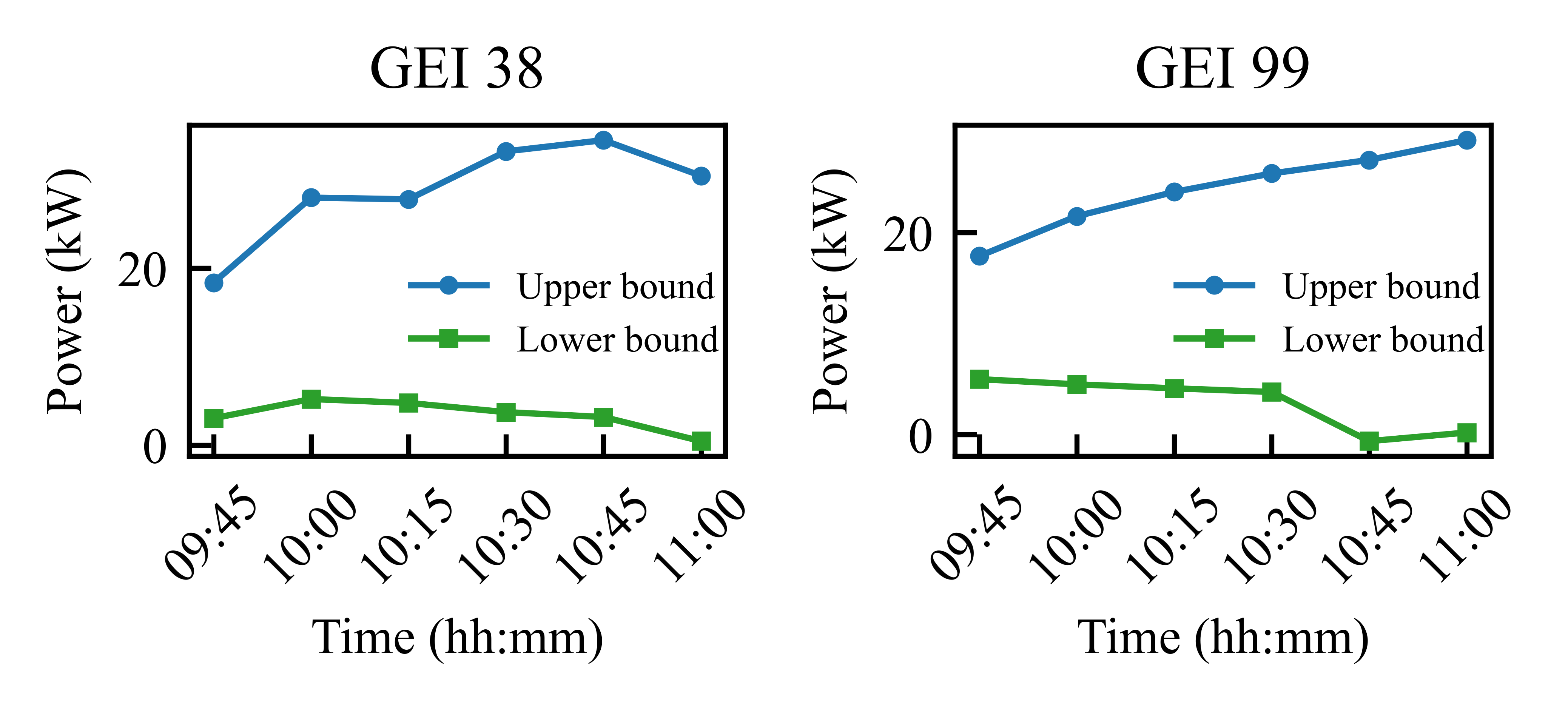}
     \caption{Flexibility range estimates of GEI 38 and 99 at 9:45.}
     \label{flexibility_range}
 \end{figure}

\begin{figure}[th]
     \centering
     \includegraphics[scale=1]{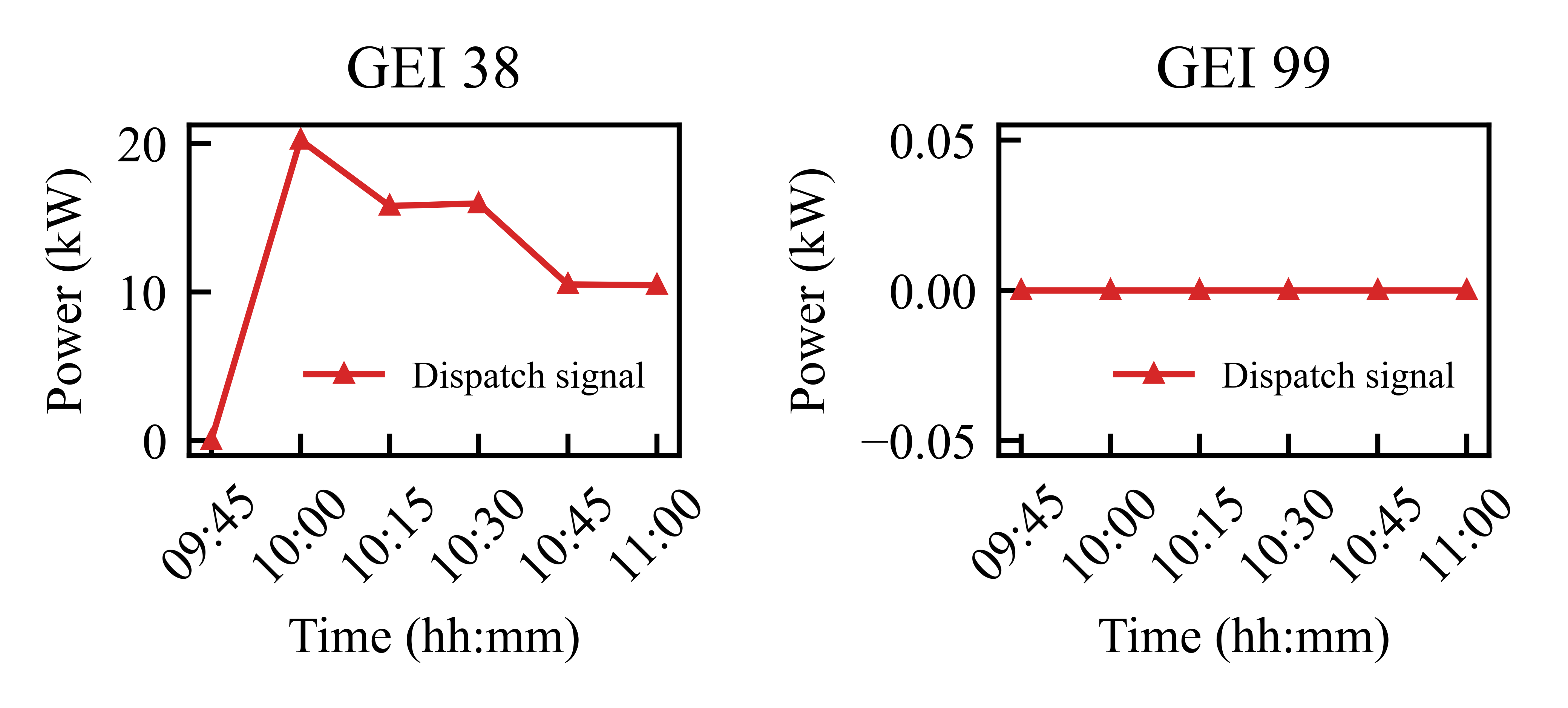}
     \caption{Dispatch signal communicated to GEI 38 and 99 at 9:45.}
     \label{dispatch_signal}
 \end{figure}

\subsubsection{Utility and GEI interaction}
There is a continuous interaction between the utility and the GEI devices. For brevity, we will only present the interaction between GEI 38 and GEI 99 with the utility just before 9:45. 
A flexibility range along the prediction horizon from 09:45 to 11:00 communicated by GEI 38 and 99 is shown in Fig.~\ref{flexibility_range}. Utility utilizes these flexibility ranges and provides a dispatch signal across the prediction horizon to each GEI, out of which the dispatch signal for GEI 38 and 99 is shown in Fig. \ref{dispatch_signal}. Since GEI 99 is not yet connected to the utility at 9:45, the dispatch signal is zero.  
\subsubsection{GEI-level analysis}
\begin{figure}[t]
     \centering
     \includegraphics[scale=1]{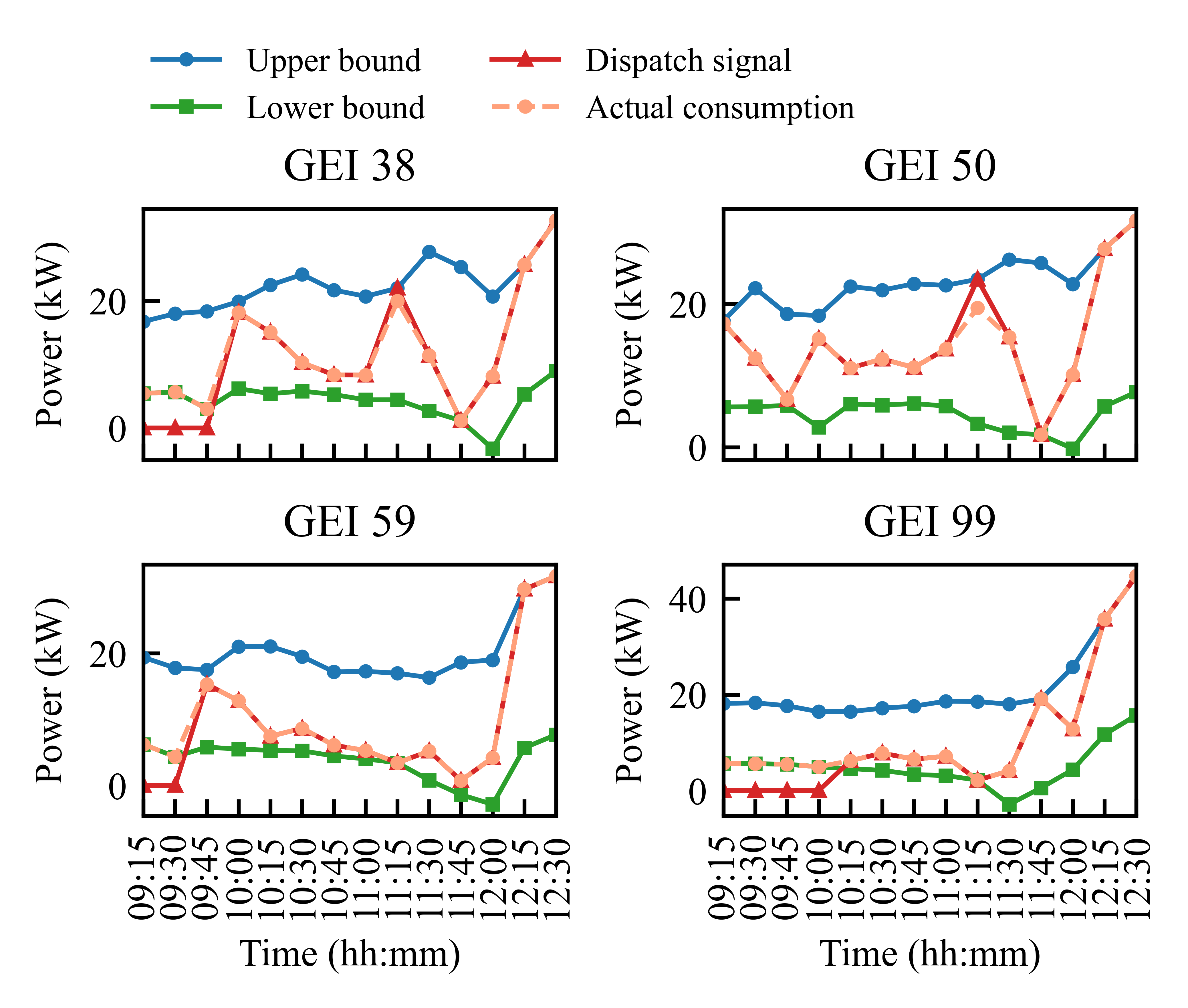}
     \caption{Coordinated operation of GEI 38,50, 59, and 90 with the utility during BS and restoration period.}
     \label{Schedule of GEI}
 \end{figure}
The GEI is designed for standalone operation during a utility blackout, operating at a minimum power level to ensure longevity. Once connected to the utility, it begins to track the dispatch signal. In addition, it keeps on communicating the flexibility range at every time step. In this way, the GEI serves a dual purpose: (1) supporting the utility in the BS process through flexible operation, and (2) meeting its own load requirements. 

The high-level operation of four randomly selected GEI devices -- GEI 38, GEI 50, GEI 59, and GEI 99 -- is shown in Fig. \ref{Schedule of GEI}. GEI 38 and GEI 59 operate close to the lower bound of their flexibility range before connecting to the utility, as observed prior to 10:00 and 9:45, respectively. In the same time frame, the dispatch signal is zero. Once connected to the utility, they closely follow the dispatch signal, represented by the red line in Fig. \ref{Schedule of GEI}. It is important to note that the dispatch signal always remains within the GEI's flexibility range due to coordinated operation between the utility and the GEI. Similarly, GEI 50 and GEI 99 are already connected to the utility by 9:15 and continue to track the dispatch signal throughout the BS process.

\begin{figure}[th]
     \centering
     \includegraphics[scale=1]{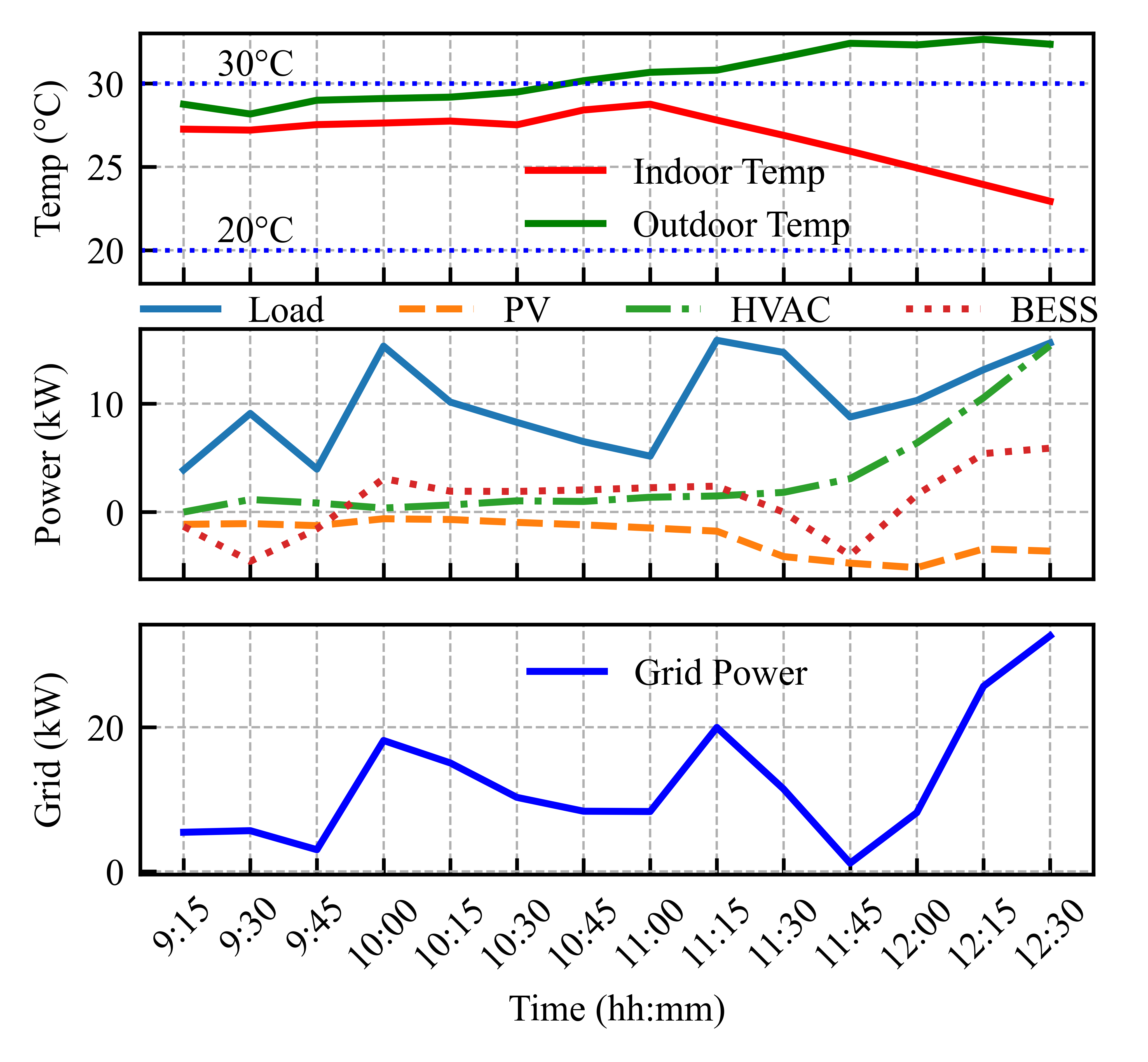}
     \caption{Operation of house 38 with GEI during the restoration period.}
     \label{GEI38}
 \end{figure}

Then, GEI 38 is selected for further analysis of its performance throughout the entire BS process. As shown in Fig.~\ref{GEI38}, GEI 38 minimizes its load and HVAC consumption while maximizing BESS discharge and PV generation until it is restored at 10:00. Once restored, its power consumption and generation follow the utility's dispatch signal. Throughout the process, the indoor temperature remains within the acceptable comfort range, demonstrating effective coordination of energy resources.

\subsection{Comparison with different flexible load level }
\begin{figure}[th]
     \centering
     \includegraphics[scale=1]{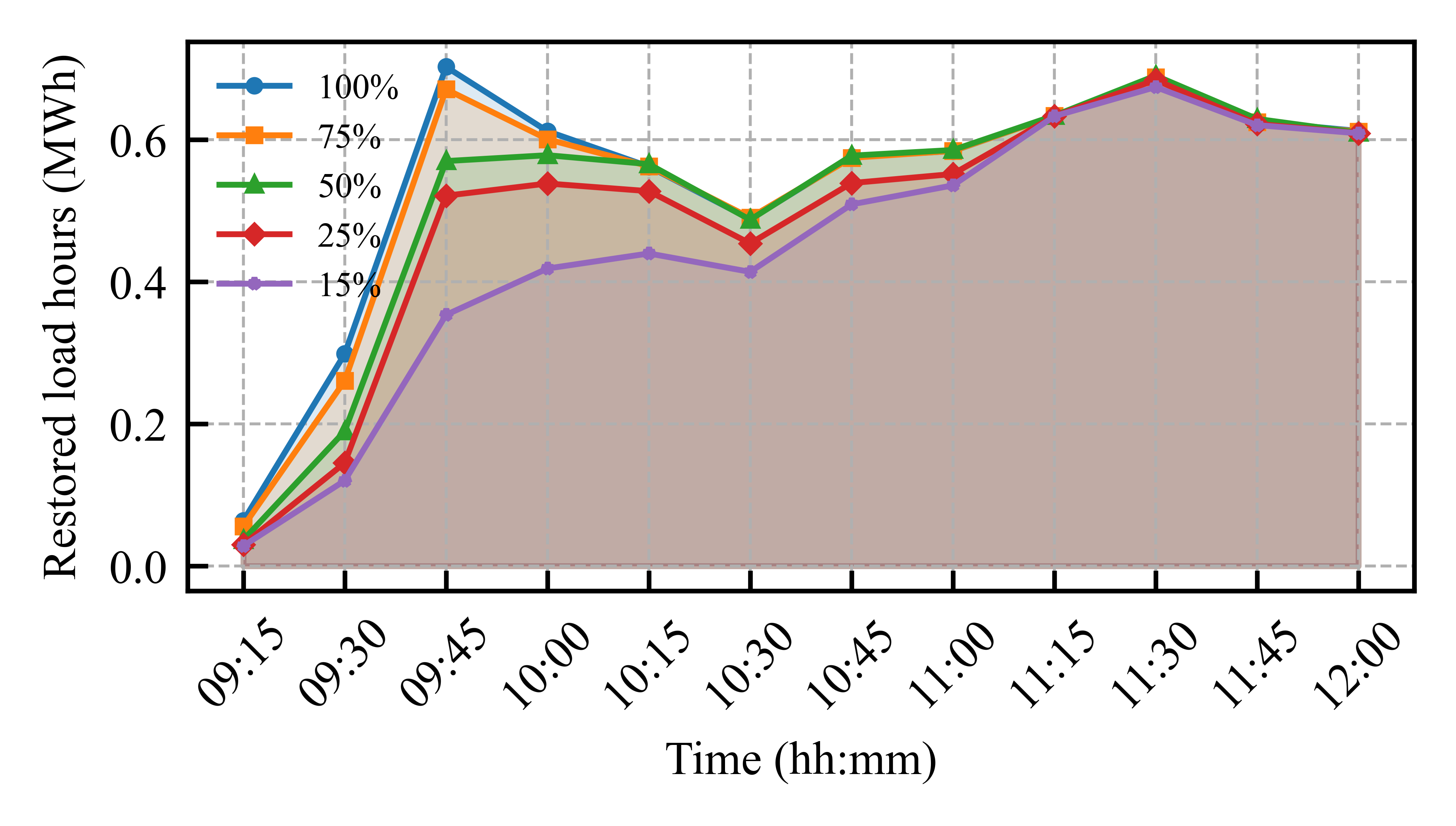}
     \caption{Comparison in terms of restored load hours.}
     \label{Restored load}
 \end{figure}

\begin{figure}[th]
     \centering
     \includegraphics[scale=1]{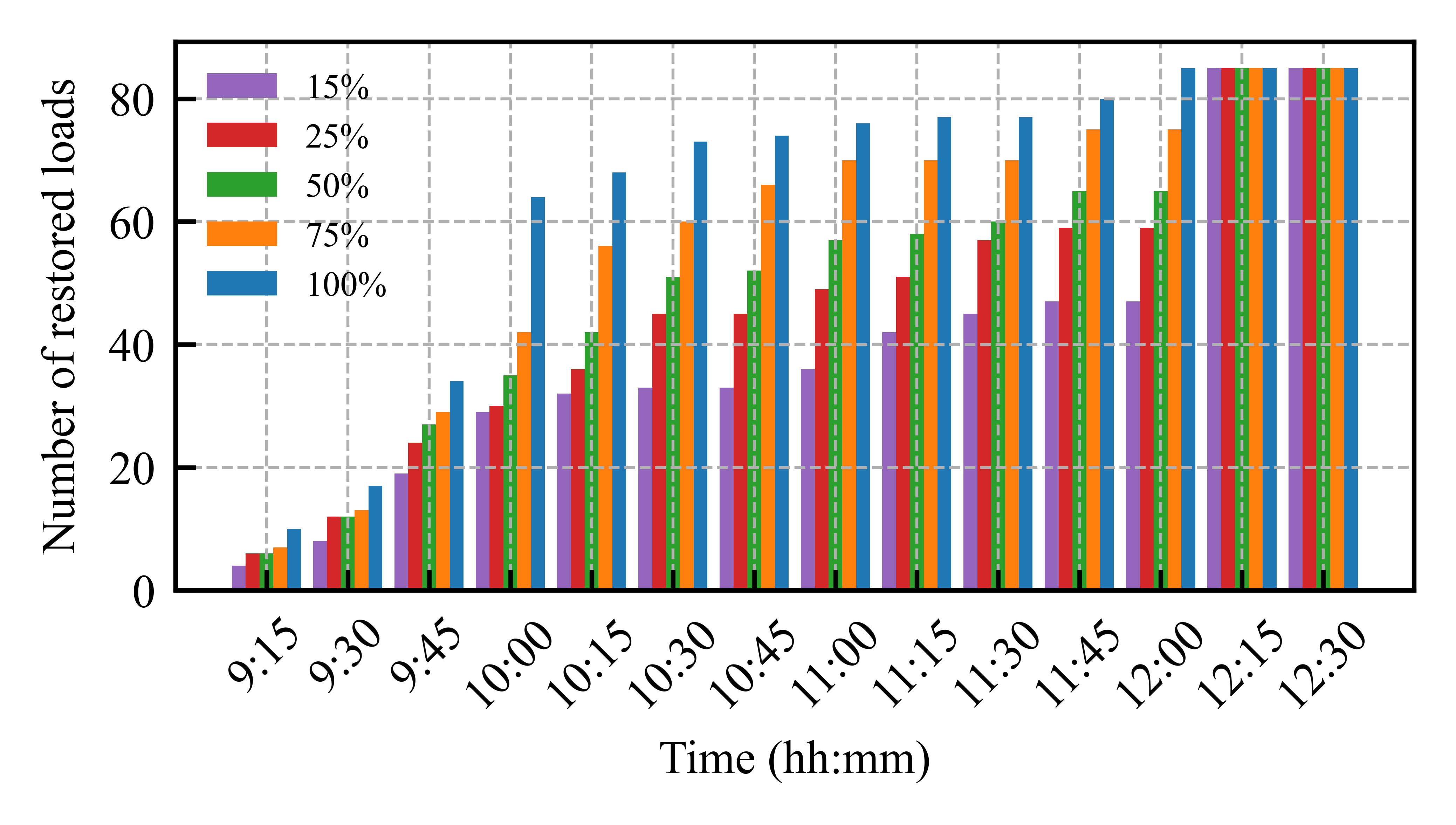}
     \caption{Comparison in terms of number of houses being restored.}
     \label{Load number}
 \end{figure}

To demonstrate the advantages of GEI during BS and load restoration, several simulations are conducted with varying levels of GEI penetration. In this study, residential houses are progressively equipped with GEI devices, ranging from 15\% to 100\%, and the proposed framework is evaluated for each case. The final simulation results are compared in terms of restored load-hours and the number of restored loads, as shown in Fig. \ref{Restored load} and Fig. \ref{Load number}, respectively.

With only 15\% GEI penetration, the total restored load-hours amount to 5.36 MWh, whereas at 100\% penetration, it increases to 6.44 MWh. As shown in Fig.~\ref{Load number}, the number of houses restored is consistently higher with 100\% GEI at each time step prior to TG synchronization, which occurs at 12:00. For other penetration levels, significant number of houses are restored after TG synchronization. These results clearly indicate that higher GEI penetration not only accelerates the restoration process but also enables the system to serve a significant portion of energy demand during the outage period.


\section{Conclusion}
This paper proposes a predictive framework for a bottom-up black start strategy that leverages dynamic coordination between GEI and the utility to accelerate load restoration in the DS. The proposed coordination scheme requires GEI devices to communicate their flexibility ranges and the utility to provide dispatch signals, without the need to share detailed asset information between the two parties. A predictive model for GEI is developed to accurately quantify multi-period flexibility ranges and to track the utility’s dispatch signals. Furthermore, the proposed BS strategy incorporates frequency security constraints, energization conditions, and synchronization requirements, enabling the dynamic formation of multiple microgrids and their integration to facilitate mutual sharing of power and energy capacity. The effectiveness of the proposed method is validated on a modified IEEE 123-bus test system, configured with two GFMIs and evaluated under varying levels of GEI penetration. The results demonstrate that the approach significantly enhances system resilience by enabling rapid load restoration and maximizing the utilization of behind-the-meter DERs with GEI.


%

\appendices




\ifCLASSOPTIONcaptionsoff
  \newpage
\fi


\begin{thebibliography}{99}

\bibitem{Wang2015}
Z.~Wang and J.~Wang, ``Self-healing resilient distribution systems based on sectionalization into microgrids,'' \textit{IEEE Trans. Power Syst.}, vol.~30, no.~6, pp.~3139--3149, Nov.~2015.

\bibitem{Wang2016}
Z.~Wang, B.~Chen, J.~Wang, and C.~Chen, ``Networked microgrids for self-healing power systems,'' \textit{IEEE Trans. Smart Grid}, vol.~7, no.~1, pp.~310--319, Jan.~2016.

\bibitem{Sun2011}
W.~Sun, C.-C.~Liu, and L.~Zhang, ``Optimal Generator Start-Up Strategy for Bulk Power System Restoration,'' \textit{IEEE Trans. Power Syst.}, vol.~26, no.~3, pp.~1357--1366, Aug.~2011.

\bibitem{Patsakis2018}
G.~Patsakis, D.~Rajan, I.~Aravena, J.~Rios, and S.~Oren, ``Optimal Black Start Allocation for Power System Restoration,'' \textit{IEEE Trans. Power Syst.}, vol.~33, no.~6, pp.~6766--6776, Nov.~2018.

\bibitem{Wang2025}
Q.~Wang, X.~Zhang, Y.~Xu, Z.~Yi, and D.~Xu, ``Planning of stationary-mobile integrated battery energy storage systems under severe convective weather,'' \textit{IEEE Trans. Sustain. Energy}, vol.~16, no.~2, pp.~1253--1268, Apr.~2025.

\bibitem{Braun2018}
M.~Braun \textit{et al.}, ``The future of power system restoration: Using distributed energy resources as a force to get back online,'' \textit{IEEE Power Energy Mag.}, vol.~16, no.~6, pp.~30--41, Nov.--Dec.~2018.

\bibitem{liu2021}
W.~Liu and F.~Ding, ``Hierarchical Distribution System Adaptive Restoration With Diverse Distributed Energy Resources,'' \textit{IEEE Trans. Sustain. Energy}, vol.~12, no.~2, pp.~1347--1359, Apr.~2021.

\bibitem{wang2019}
Y.~Wang \textit{et al.}, ``Coordinating Multiple Sources for Service Restoration to Enhance Resilience of Distribution Systems,'' \textit{IEEE Trans. Smart Grid}, vol.~10, no.~5, pp.~5781--5793, Sep.~2019.

\bibitem{sadeque2023}
F.~Sadeque and B.~Mirafzal, ``Frequency Restoration of Grid-Forming Inverters in Pulse Load and Plug-in Events,'' \textit{IEEE J. Emerg. Sel. Topics Ind. Electron.}, vol.~4, no.~2, pp.~580--588, Apr.~2023.

\bibitem{Maharjan2025}
S.~Maharjan, C.~Bai, H.~Wang, Y.~Yao, F.~Ding, and Z.~Wang, ``Distribution System Blackstart and Restoration Using DERs and Dynamically Formed Microgrids,'' \textit{IEEE Trans. Smart Grid}, 2025.

\bibitem{Maharjan2023}
S.~Maharjan, A.~Trivedi, and D.~Srinivasan, ``Rules-integrated model predictive control of office space for optimal electricity prosumption,'' \textit{Sustain. Energy Grids Netw.}, vol.~33, pp.~100981, 2023.

\bibitem{Arif2018}
A.~Arif, S.~Ma, Z.~Wang, J.~Wang, S.~M.~Ryan, and C.~Chen, ``Optimizing service restoration in distribution systems with uncertain repair time and demand,'' \textit{IEEE Trans. Power Syst.}, vol.~33, no.~6, pp.~6828--6838, Nov.~2018.

\bibitem{Chen2019}
B.~Chen, Z.~Ye, C.~Chen, and J.~Wang, ``Toward a MILP modeling framework for distribution system restoration,'' \textit{IEEE Trans. Power Syst.}, vol.~34, no.~3, pp.~1749--1760, May~2019.

\bibitem{Arif2022}
A.~Arif, B.~Cui, and Z.~Wang, ``Switching device-cognizant sequential distribution system restoration,'' \textit{IEEE Trans. Power Syst.}, vol.~37, no.~1, pp.~317--329, Jan.~2022.

\bibitem{Bedoya2021}
J.~C.~Bedoya, Y.~Wang, and C.-C.~Liu, ``Distribution system resilience under asynchronous information using deep reinforcement learning,'' \textit{IEEE Trans. Power Syst.}, vol.~36, no.~5, pp.~4235--4245, Sept.~2021.

\bibitem{Du2022}
Y.~Du and D.~Wu, ``Deep reinforcement learning from demonstrations to assist service restoration in islanded microgrids,'' \textit{IEEE Trans. Sustain. Energy}, vol.~13, no.~2, pp.~1062--1072, Apr.~2022.

\bibitem{bassey2019}
O.~Bassey, K.~L.~Butler-Purry, and B.~Chen, ``Dynamic Modeling of Sequential Service Restoration in Islanded Single Master Microgrids,'' \textit{IEEE Trans. Power Syst.}, vol.~35, no.~1, pp.~202--214, Jan.~2020.

\bibitem{che2019}
L.~Che and M.~Shahidehpour, ``Adaptive Formation of Microgrids With Mobile Emergency Resources for Critical Service Restoration in Extreme Conditions,'' \textit{IEEE Trans. Power Syst.}, vol.~34, no.~1, pp.~742--753, Jan.~2019.

\bibitem{zhang2021}
Q.~Zhang, Z.~Ma, Y.~Zhu, and Z.~Wang, ``A Two-Level Simulation-Assisted Sequential Distribution System Restoration Model With Frequency Dynamics Constraints,'' \textit{IEEE Trans. Smart Grid}, vol.~12, no.~5, pp.~3835--3846, Sept.~2021.

\bibitem{song2021}
M.~Song, R.~R.~Nejad, and W.~Sun, ``Robust Distribution System Load Restoration With Time-Dependent Cold Load Pickup,'' \textit{IEEE Trans. Power Syst.}, vol.~36, no.~4, pp.~3204--3215, Jul.~2021.

\bibitem{taha2019}
A.~F.~Taha, N.~Gatsis, B.~Dong, A.~Pipri, and Z.~Li, ``Buildings-to-grid integration framework,'' \textit{IEEE Trans. Smart Grid}, vol.~10, no.~2, pp.~1237--1249, Mar.~2019.

\bibitem{jin2025}
X.~Jin, X.~Wang, H.~Jia, Y.~Mu, Q.-W.~Wu, and W.~Wei, ``Peer-to-peer multi-energy trading among heterogeneous building prosumers via asynchronous distributed algorithm,'' \textit{IEEE Trans. Smart Grid}, vol.~16, no.~2, pp.~1590--1603, Mar.~2025.

\bibitem{cheng2022}
R.~Cheng, Z.~Wang, Y.~Guo, and Q.~Zhang, ``Online voltage control for unbalanced distribution networks using projected Newton method,'' \textit{IEEE Trans. Power Syst.}, vol.~37, no.~6, pp.~4747--4760, Nov.~2022.

\end{thebibliography}
\end{document}